\documentclass[pre,aps,twocolumn,showpacs,amsmath,amssymb,superscriptaddress]{revtex4-1}
\usepackage{amsmath}
\usepackage{amssymb}
\usepackage{graphicx}
\usepackage{euscript}

\newlength{\pictwidth}
\begin{document}
\title{Study of wave chaos in a randomly-inhomogeneous oceanic acoustic waveguide:
spectral analysis of the finite-range evolution operator}

\author{D.V. Makarov}\email{makarov@poi.dvo.ru}

\author{L.E. Kon'kov}

\author{M.Yu. Uleysky}
\affiliation{V.I.Il'ichev Pacific Oceanological Institute \\
of the Far-Eastern Branch of the Russian
Academy of Sciences, \\ 43 Baltiyskaya St.,
690041, Vladivostok, Russia}

\author{P.S. Petrov}
\affiliation{V.I.Il'ichev Pacific Oceanological Institute \\
of the Far-Eastern Branch of the Russian
Academy of Sciences, \\ 43 Baltiyskaya St.,
690041, Vladivostok, Russia}
\affiliation{Far-East Federal University, \\ 
8 Sukhanova St., 690950, Vladivostok, Russia}

\begin{abstract}
The proplem of sound propagation
in an oceanic waveguide is considered.
 Scattering on random inhomogeneity of the waveguide
leads to wave chaos.
Chaos reveals itself in spectral properties
of the finite-range evolution operator (FREO). FREO describes transformation
of a wavefield in course of propagation along a finite segment of a waveguide.
We study transition to chaos by 
tracking variations in spectral statistics with increasing length of the segment.
Analysis of the FREO is accompanied with ray calculations
using the one-step Poincar\'e map which is the classical counterpart of the FREO.
Underwater sound channel in the Sea of Japan is taken
for an example.
Several methods of spectral analysis are utilized.
In particular, we approximate level spacing statistics by means of the Berry-Robnik
and Brody distributions, explore the spectrum using the procedure
elaborated by A.~Relano with coworkers (Relano et al, Phys. Rev. Lett., 2002; Relano, Phys. Rev. Lett., 2008),
and analyze modal expansions of the eigenfunctions.
We show that the analysis of FREO eigenfunctions is more informative than the analysis
of eigenvalue statistics.
It is found that near-axial sound propagation in the Sea of Japan
preserves stability even over distances of hundreds kilometers.
This phenomenon is associated
with the presence of a shearless torus in the classical 
phase space.
Increasing of acoustic wavelength degrades scattering, resulting in recovery of localization
near periodic orbits of the one-step Poincar\'e map.
Relying upon the formal analogy between wave and quantum chaos,
we suggest that the concept of FREO, supported by classical calculations via the one-step Poincar\'e map, 
can be efficiently applied for studying chaos-induced decoherence in quantum systems.

\end{abstract}
\maketitle

\section{Introduction}

Sound speed in the deep ocean typically has a minimum at some depth.
This results in formation of a refractive waveguide, the so-called underwater
sound channel, which prevents sound waves from contact with the absorbing bottom.
As sound absorption within water column is fairly weak, an underwater
sound channel enables sound propagation over distances of thousands
kilometers.
The largest distance had been achieved using explosive charges
in the seminal experiment 
on sound transmission from Perth to Bermuda in 1960 \cite{Shockley,MunkRR}.

 It is realized that small sound-speed variations
induced by oceanic internal waves lead to Lyapunov instability and chaos
of sound rays.
In the mathematical sense,
ray chaos is an analogue of classical chaos
in Hamiltonian systems.
Indeed, ray motion in a waveguide is equivalent to
motion of a point particle in a potential well,
and sound-speed variations along a waveguide
play the role of a nonstationary perturbation.
Reciprocal Lyapunov exponent for chaotic rays typically is of
about several tens kilometers \cite{RayWave}, therefore, the problem
of ray chaos is mainly important for long-range sound propagation.
During the last two decades ray chaos in ocean acoustics
was an object of intense research, both theoretical and experimental \cite{Smith1,Viro01,Review03,Chaos,BV04}.
Considerable attention was paid to wave chaos  \cite{RayWave,Viro99,Viro04,Viro05,PRE76,Acoust08,UFN}. 
The term wave chaos
relates to wavefield manifestations of ray chaos.
It was found that interference makes wave refraction more regular than it is anticipated
from ray modeling, albeit influence of ray chaos
persist even for very low sound frequencies \cite{Hege}.
This problem becomes especially important due to the growing interest
to hydroacoustical tomography, i.~e.
monitoring of environment using sound signals. The classical scheme 
of tomography developed by Munk and Wunsch \cite{MW79} is based on computation
of eigenrays connecting the source and the receiver.
It was shown in \cite{Tap} that ray chaos leads to exponential proliferation
of eigenrays with increasing distance. As a result, the inverse problem
becomes ill-posed, impeding environment reconstruction.
However, wave-based corrections ``stabilize'' wave refraction,
i.~e. the standard semiclassical approximation typically overestimates ray chaos.
Thus, one needs either an improved version of the semiclassical approximation for proper computation
of eigenrays, or some approach for making implications about eigenray stability relying upon
wave modeling, whereby a priori taking into account the wave-based suppression of ray chaos.

Theory of ray and wave chaos extensively 
exploits methods borrowed from the theory of 
Hamiltonian dynamical systems \cite{RayWave,Zas}.
In particular, ray motion is often studied by means of 
the phase space representation. This provides
the clear geometric interpretation of ray dynamics 
for toy models of deterministic range-periodic waveguides.
For instance, one can easily separate domains of initial 
conditions corresponding to stable and chaotic rays from each other.
However, realistic underwater sound channels are not range-periodic,
and their range inhomogeneity should be rather described as a stochastic
process. 
On the other hand, statistical methods \cite{DozierI,ColosiMorozov}
are implicitly based on the assumption of ergodic chaos and
ignore the existence 
of phase space domains of finite-range stability \cite{WT01,PRE73}.
Therefore, it is reasonable to elaborate some theoretical approach
combining advantages of statistical and deterministic methods. 
Recently it had been shown that a bridge
between the deterministic and statistical descriptions
of wave propagation in random media can be built up 
by introducing the finite-range evolution operator 
(hereafter FREO) \cite{Arxiv,UFN}.
FREO acts as a propagator determining wavefield transformation
between two vertical sections of a waveguide.
Spectral analysis of the FREO
in terms of the random matrix theory
allows one to estimate contribution of ray chaos
to wave dynamics and track transition to chaos
with increasing distance between the sections.
A somewhat different approach had been offered in
\cite{Tomc11}. 
These approaches allow one to generalize the well-developed
spectral theory of quantum chaos (see, for instance, \cite{Stockman}) 
on one-way wave propagation in random media.
On the other hand, concept of the FREO can serve as a promising
method for studying 
chaos-induced decoherence in randomly-driven
quantum systems \cite{Kol97}.

In the present paper we study spectral statistics of the FREO
for the Sea of Japan.
Attention is concentrated on the track connecting
the Gamov peninsula and Kita-Yamato bank.
The length of the track is about 350 km. 
Our interest to this waveguide is motivated by results
of the experiment conducted there in 2006 \cite{Bez09},
indicating on high stability of near-axial propagation.
Similar behavior was observed in a earlier experiment
with a slightly different propagation track \cite{IEEE}.
It should be noted that
stability of near-axial propagation is atypical for the deep ocean \cite{RayWave,UFN}.

The paper is organized as follows.
The next section represents basic equations describing 
long-range sound propagation in the ocean. 
In Section \ref{Model}, we describe
the waveguide used in the paper.
Section \ref{Operator} contains description of the FREO and methods
of its analysis. Section \ref{Onestep} is devoted to
the classical counterpart of the FREO, namely the one-step Poincar\'e map.
In Section \ref{Japan}, we perform statistical analysis
of the FREO for the underwater sound channel in the Sea of Japan.
In Conclusion we summarize and discuss the results obtained.

\section{General equations}
\label{General}

Ocean is a layered media, and its horizontal variability
is much weaker than vertical one.
This allows one to reduce the initial three-dimensional
problem of wave propagation in an underwater sound channel
to a two-dimensional one by
assuming cylindrical symmetry and
neglecting azimuthal coupling.
Sound refraction is governed by spatial variability
of sound speed
\begin{equation}
 c(z,\,r)=c_0+\Delta c(z)+\delta c(z,\,r),
\label{c}
\end{equation}
where $z$ is depth, $r$ is range coordinate,
$c_0$ is a reference sound speed.
Sound-speed variations
obey the double inequality
\begin{equation}
\lvert\delta c\rvert_\text{max}\ll \lvert\Delta c\rvert_\text{max}\ll c_0.
\label{ocean}
\end{equation}
Left inequality implies that the range-dependent term 
can be treated as a weak perturbation of 
the background sound-speed profile.
This term is mainly contributed from oceanic internal waves.
Right inequality means that variations of the refractive index
are weak, and only those waves which propagate with small angles
with respect to the horizontal plane can avoid contact with
the absorbing ocean bottom.
Thus, one can invoke the small-angle approximation, in which
an acoustic wavefield is governed by the standard parabolic
equation
\begin{equation}
\frac{i}{k_0}\frac{\partial\Phi}{\partial r}=
-\frac{1}{2k_0^2}\frac{\partial^2\Phi}{\partial
z^2}+\left[U(z)+\varepsilon V(z,\,r)\right]\Phi, 
\label{parabolic}
\end{equation}
where 
wave function $\Phi$ is linked to
acoustic pressure $u$ by means of the formula $u=\Phi\exp(ik_0r)/\sqrt{r}$. 
Here the denominator $\sqrt{r}$ responds for the cylindrical spreading of sound.
Quantity $k_0$ is the reference wavenumber related to sound
frequency $f$ as $k_0=2\pi f/c_0$.
Functions $U(z)$ and $V(z,r)$ are determined by spatial sound-speed variations.
In the the small-angle approximation they can be expressed as
\begin{equation}
U(z)=\frac{\Delta c(z)}{c_0},\quad 
V(z,\,r)=\frac{\delta c(z,\,r)}{c_0}. 
\label{pot}
\end{equation}
According to (\ref{ocean}), 
$\lvert V\rvert_\text{max}\ll\lvert U\rvert_\text{max}$, that is,
function $V(z,r)$ can be treated as a small perturbation.
One can easily see that the replacement
\begin{equation}
 k_0^{-1}\to\hbar,\quad
r\to t
\end{equation}
transforms the parabolic equation (\ref{parabolic}) into the Schr\"odinger equation
for a particle with unit mass.
This circumstance enables study of wave propagation
using the approaches developed in quantum mechanics.
In this relationship, function $U(z)$ serves as an unperturbed potential.
As $r$ is a timelike variable, $V(z,\,r)$ plays the role of a nonstationary
perturbation.

In the short-wavelength limit $k_0\to\infty$, 
solution of the parabolic equation (\ref{parabolic})
can be expressed as a sum of rays whose
trajectories are governed
by the Hamiltonian
\begin{equation}
H=\frac{p^2}{2}+U(z)+V(z,\,r),
 \label{hamilt}
\end{equation}
where $p=\tan\chi$, $\chi$ is ray grazing angle (i.~e. angle with respect
to the horizontal plane).
The respective Hamiltonian equations read
\begin{equation}
\frac{dz}{dr}=\frac{\partial H}{\partial p}=p,\quad 
\frac{dp}{dr}=-\frac{\partial H}{\partial z}= 
-\frac{\partial U}{\partial z}-\frac{\partial V}{\partial z}.
\label{sys}
\end{equation}
Owing to the analogy with classical mechanics,
$p$  is referred to as {\it ray momentum}.

\section{Model of a waveguide}
\label{Model}

\subsection{Background sound-speed profile}
\label{Background}

\begin{figure}[!htb]
\begin{center}
\includegraphics[width=0.48\textwidth]{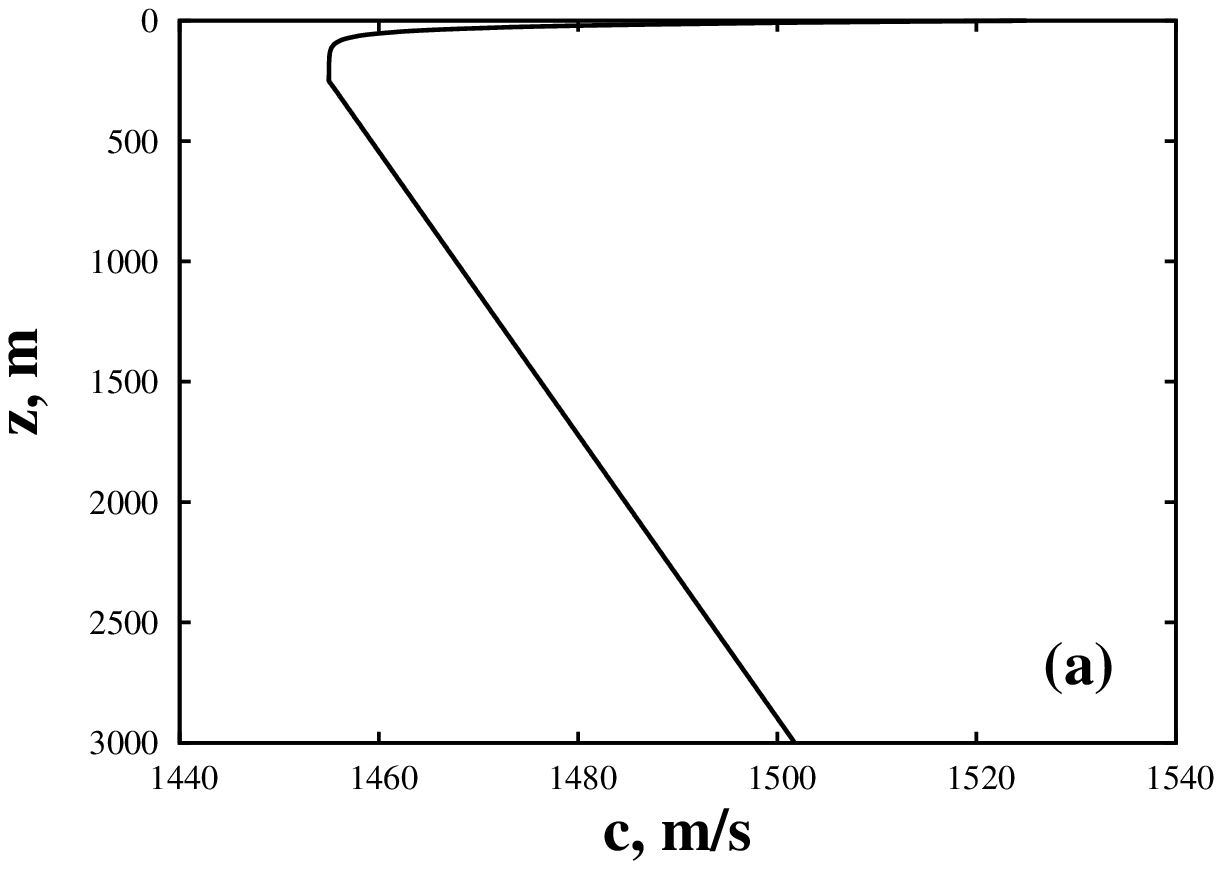}\\
\includegraphics[width=0.48\textwidth]{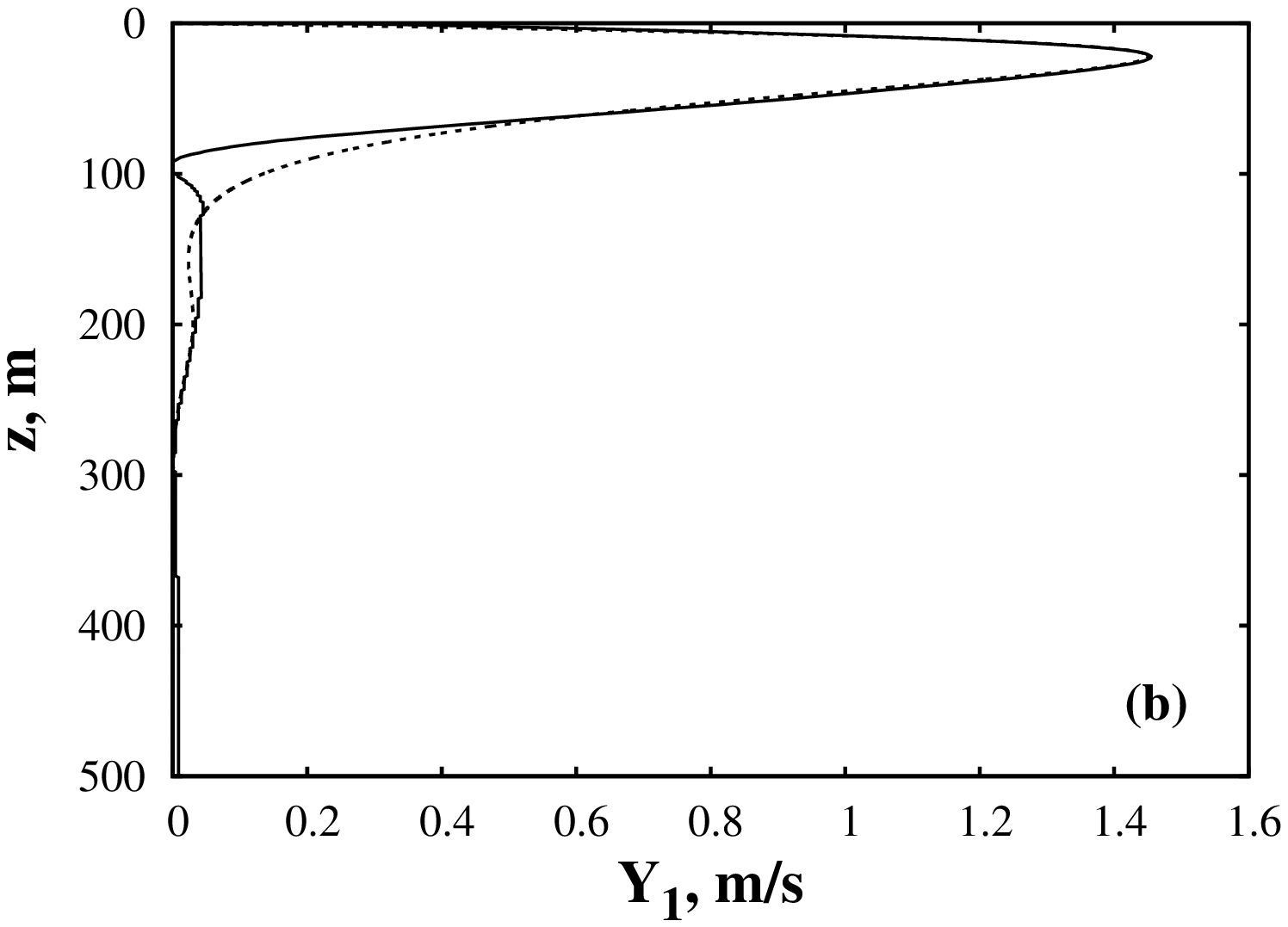}
\caption{
(a) Unperturbed sound-speed profile,
(b) the first empirical orthogonal function of the sound-speed
perturbation (solid) and its smoothed approximation 
(dotted).
}%
\label{fig-soundspeed}
\end{center}
\end{figure}

Model of the underwater sound channel in the Sea of Japan
was elaborated using the hydrological data from
the database \cite{Database}.
Function $U(z)$ corresponding to the background sound-speed profile was approximated
by the expression
%
\begin{equation}
U(z)=\left\{
\begin{aligned}
&U_1(z), &z\leqslant z_0,\\
&U_2(z), &z> z_0.
\end{aligned}\right.
\label{uz}
\end{equation}
where
\begin{equation}
\begin{aligned}
 U_1(z)&=\frac{c_1}{c_0}e^{-z/z_1},\\
 U_2(z)&=\frac{c_1}{c_0}e^{-z_0/z_1}+\frac{g}{c_0}(z-z_0),
\end{aligned}
\label{uz12}
\end{equation}
$c_0=1455$~m/s, $c_1=70$~m/s, $z_0=250$~m is the depth of the channel axis, i.~e.
the depth with the minimal sound speed,
$z_1=30$~m, $g=0.017$~s$^{-1}$.
 (see Fig.~\ref{fig-soundspeed}).
The ocean bottom is assumed to be flat and placed at the depth $h=3$ km.
We consider only the deep-water propagation, albeit
the source in the aforementioned experiments \cite{Bez09,IEEE} was mounted into the bottom
in the coastal zone near the Gamov peninsula. 
The shallow-water part of the waveguide was relatively short,
less than 30 km, and didn't have significant bottom features which
could remarkably alter ray stability.

\begin{figure}[!htb]
\includegraphics[width=0.48\textwidth]{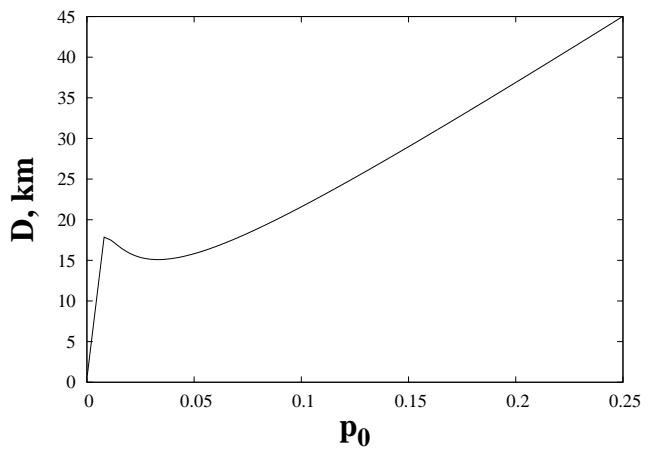}
\caption{
Ray cycle length vs initial ray momentum for the source located at the channel axis.
}%
\label{fig-dp}
\end{figure}

Expressions (\ref{uz}) and (\ref{uz12}) permit analytical derivation
of the basic model characteritics in the absence of horizontal
inhomogeneity.
One can introduce the ray action 
\begin{equation}
 I = \frac{1}{\pi}\int\limits_{z_\text{min}}^{z_\text{max}}\sqrt{2[E-U(z)]}\,dz,
\label{action}
\end{equation}
where $z_\text{min}$ and $z_\text{max}$ are the upper and lower
ray turning points, respectively,
and $E=0.5p^2 + U(z)$.
Ray action describes steepness of a ray trajectory
and enters into the Einstein-Brillouin-Keller
quantization rule.
\begin{equation}
k_0I_m = m - 1/2,\quad 
m = 1,2,\dots
\label{EBK}
\end{equation}
establishing
the link between normal modes of the unperturbed waveguide 
and modal rays.
Here $I_m$ is the action of a modal ray,
and both ray turning points are assumed to be inside
the water column,
that is, rays undergo total internal reflection due to
smooth vertical gradient of the refractive index $n = c_0/c(z,\,r)$.
Normal modes are the solutions of the Sturm-Liouville equation
\begin{equation}
-\frac{1}{2k_0^2}\frac{\partial^2\phi_n(z)}{\partial
 z^2}+U(z)\phi_n(z)=E_n\phi_n(z),
\label{StL}
\end{equation}
with appropriate boundary conditions at the ocean surface and bottom.
The main contribution to the mode $m$ is given from the rays
with $I\simeq I_m$ \cite{Viro99}.
Integration of (\ref{action}) yields
\begin{multline}
I = 2z_1\sqrt{2E_\text{min}}
\left[\sqrt{\varepsilon}\ln(\sqrt{\varepsilon}+\sqrt{\varepsilon-1})-\sqrt{\varepsilon-1}\vphantom{\sqrt{\mathstrut}}\right]+\\
\frac{2\sqrt{2}c_0}{3g}(E-E_\text{min})^{3/2},
\label{action1}
\end{multline}
where $\varepsilon = E/E_\text{min}$, $E_\text{min}=(c_1/c_0)\exp(-z_0/z_1)$.
Ray cycle length, i.~e. horizontal distance
between two successive upper (or lower) ray turning points,
can be determined as
\begin{multline}
D(E) = 2\pi\frac{dI}{dE} = \\
2z_1\sqrt{\frac{2}{E}}\ln\left(\sqrt{\varepsilon} + \sqrt{\varepsilon- 1}\right)
+ \frac{2c_0\sqrt{2E_\text{min}}}{g}\sqrt{\varepsilon - 1}.
\label{dIdE}
\end{multline}
Fig.~\ref{fig-dp} represents dependence of ray cycle length
on initial ray momentum for the source located at the channel axis $z=250$~m.
Initial momentum $p_0$ depends on $E$ as
$p_0=\sqrt{2(E-E_\text{min})}$. 
Function $D(p_0)$ is nonmonotonic and has two extrema,
the sharp maximum and the smooth minimum.
The latter one can give rise to a so-called weakly-divergent beam
 \cite{Brekh90,Caruth,MorCol,Petukhov}. 
Its low divergence is associated with approximate equality
of cycle length values for rays forming the beam.
It will be demonstrated in Section \ref{Onestep} that 
the local minimum of $D(p_0)$
plays a significant role in ray stabilty.

\subsection{Sound-speed perturbation}
\label{Perturbation}

Model of the internal-wave-induced
sound-speed perturbation was built up in several steps.
Firstly, we computed the range-averaged profile
of buoyancy frequency, using the hydrological data
\cite{Database}.
Then, we calculated  realizations
of the sound-speed perturbation using the method proposed in \cite{CB}.
 In order to facilitate numerical simulation,
the perturbation was expanded over
empirical orthogonal functions
\cite{LeBlanc}
\begin{equation}
 \delta c(z,\,r)=\left<\delta c(z)\right>+\sum\limits_n b_n(r)Y_n(z).
\label{KL}
\end{equation}
Empirical orthogonal functions  $Y_n(z)$ 
are the eigenvectors of the covariance matrix
 $\hat K$ with elements
\begin{equation}
K_{ij}=\frac{1}{L}\sum\limits_{l=1}^{L}[\delta c_l(z_i)-\left<{\delta c}(z_i)\right>][\delta c_l(z_j)-\left<{\delta c}(z_j)\right>],
 \label{covar}
\end{equation}
where index $l$ numbers $L$ statistically independent realizations of $\delta c(z)$, 
$\{z_i\}$ is a vector of depth values, 
and angular brackets mean ensemble average.
As  $\delta c$ is caused by internal waves, 
one can set $\left<\delta c\right>=0$.
Eigenvalues of the matrix
 $\hat K$ quantify contributions from the corresponding eigenvectors
in the expansion (\ref{KL}).
It was  found that the contribution of the first orthogonal function prevails,
and one can fairly represent the sound-speed perturbation as the product
\begin{equation}
 V(z,\,r)=b_1(r)Y_1(z)
 \label{factoriz}
\end{equation}
where $b_1(r)$ is a random function.
For simplicity, it is assumed that $b_1(r)$
is a stochastic process with 
the exponentially-decaying autocorrelation function
\begin{equation}
\left<b_1(r)b_1(r')\right>=\exp(-\lvert r-r'\rvert/\bar r),
\end{equation}
where the correlation length $\bar r$ is taken of 10~km,
that is typical for the deep ocean \cite{DozierII}.
Then the realizations of $b_1(r)$ can be computed
via the formula $b_1=\sqrt{2\bar r}\eta$, where
$\eta$ is a solution of
the Ornstein-Uhlenbeck stochastic differential equation
\begin{equation}
 \frac{d\eta}{dr}=-\frac{1}{\bar r}\eta(r) + \frac{1}{\bar r}\xi(r).
\end{equation}
The procedure of solving this equation is described, for instance, in \cite{Mallick}.
Here $\xi$ is a Gaussian white noise satisfying
\begin{equation}
\left<\xi(r)\right>=0,\quad
\left<\xi(r)\xi(r')\right>  = \delta(r-r').
\end{equation}
The resulting function 
$b_1(r)$ satisfies $\left<b_1^2(r)\right>\simeq 1$.

The calculated function  $Y_1(z)$ is depicted in Fig.~\ref{fig-soundspeed}(b).
It involves step-like changes 
 in the depth
interval from 100 to 300 meters.
These changes are caused by the depth discretization of the hydrological data
and physically irrelevant.
Wave modeling is insensitive to them, but ray-based calculations can be significantly affected.
Therefore, in ray calculations
we use a smooth analytical approximation
\begin{equation}
Y_1=Ay_{\mathrm{a}}\exp(-y_{\mathrm{a}}^n)
+B\exp(-y_{\mathrm{b}}^2),
\label{Y1} 
\end{equation}
where $A=0.0027$, $y_{\mathrm{a}}=z/z_{\mathrm{a}}$, $n=1.1$,
$z_{\mathrm{a}}=24$~m, $B=2\cdot 10^{-5}$, 
$y_{\mathrm{b}}=(z-z_{\mathrm{c}})/z_{\mathrm{b}}$,
$z_{\mathrm{b}}=50$~m, $z_{\mathrm{c}}=200$~m.

\section{Finite-range evolution operator}
\label{Operator}

Finite-range evolution operator (FREO) had been introduced 
for studying wave propagation in a randomly-inhomogeneous waveguide
in \cite{UFN,Arxiv}.
Its quantum-mechanical analogue was earlier considered in \cite{Kol97}.
Basically, a finite-range evolution operator (FREO) is an element of one-parameter group generated by the operator 
in the right-hand side of the equation (\ref{parabolic}). 
Consider a solution $\Phi(z,\,r)$ of the parabolic equation (\ref{parabolic}) complemented with the standard boundary conditions 
\begin{equation}
\left.\Phi\right\vert_{z=0} = 0,\quad
\left.\frac{d\Phi}{dz}\right\vert_{z=h}=0
\label{BCs}
\end{equation}
and the initial condition $\Phi(z,\,r=0)=\bar{\Phi}(z)$, where $\bar{\Phi}(z)$ belongs to $L^2[0,h]$ and satisfies (\ref{BCs}).
Then FREO $\hat G(\tau)$ is defined on the subspace of $L^2[0,h]$ (restricted by (\ref{BCs})) as 
\begin{equation}
 \hat G(\tau)\bar\Phi(z)\equiv\left.\Phi(z,\,r)\right\vert_{r=\tau}.
\label{evolution}
\end{equation}
By definition, the FREO describes transformation of a wavefield in course of propagation along a finite waveguide segment of length $\tau$. 
Each realization of inhomogeneity produces its own realization of the FREO.
Our interest is concerned with statistical properties of the FREO
and their connection to classical ray stability.

Note that the choice of the hard wall boundary condition at the bottom (\ref{BCs}) is typical for the deep ocean acoustics problems 
when the attention is restricted to the trapped modes, whose propagation is not affected by the bottom interaction. 
Under these conditions no energy is absorbed by the bottom. 
Hence, if the attenuation in the sea water is negligible (this is true for the sound frequencies of our interest) 
and refraction index in (\ref{parabolic}) has no imaginary part, then the FREO is a unitary operator.

FREO can be represented as a matrix in the basis of normal modes $\phi_j(z)$
satisfying the Sturm-Liouville problem (\ref{StL},\ref{BCs}).
Matrix elements of the FREO are given by  
\begin{equation}
G_{mn}(\tau)=\int\limits_0^h \phi_m(z)\hat G(\tau)\phi_n(z)\,dz.
\label{matr}
\end{equation}
Thus, the matrix elements  $G_{mn}$ are complex-valued amplitudes of 
modal transitions. For the range-independent waveguide, the matrix of FREO is diagonal with
$|G_{mm}| = 1$.

Eigenvalues and eigenvectors of the FREO obey the equation
\begin{equation}
\hat G\Psi_m(z,\,r)=g_m\Psi_m(z,\,r).
\label{eigen}
\end{equation}
Owing to unitarity, eigenvalues $g_m$ can be recast as
\begin{equation}
 g_m=e^{-ik_0\epsilon_m}, \quad 
\epsilon_m\in\Re.
\label{fm}
\end{equation}
Since eigenvalues of the FREO belong to the unit circle in the complex plane,
the FREO corresponds to the circular ensemble \cite{Stockman}.
The FREO has much in common with the Floquet operator governing wave propagation in a range-periodic
waveguide \cite{Viro05,PRE76} and quantum dynamics in time-periodic systems.
For instance, quantity $\epsilon_m$ is the analogue of quasienergy in quantum mechanics.

Note that eigenvalues of $\hat G(\tau)$ may be easily computed using its matrix representation $G_{mn}(\tau)$. 
To accomplish this one has to clip a finite block of this (infinite) matrix corresponding to the trapped modes, 
neglecting their interaction with high-order modes. This simplification is reasonable and does not affect accuracy 
of the eigenvalues computation (and the numerics confirms that) since the prevailing small-angle propagation corresponds to the low-order modes.

\subsection{Level spacing statistics}
\label{Eigv}

Wave chaos reveals itself in the spectrum of the FREO,
in particular, in the statistics of level spacings.
A level spacing is defined as
\begin{equation}
\begin{gathered}
 s=\frac{k_0M(\epsilon_{m+1}-\epsilon_m)}{2\pi},\quad 
m = 1,2,\dots,M, \\
\epsilon_{M+1} = \epsilon_1 + \frac{2\pi}{k_0}.
\end{gathered}
\label{spacing}
\end{equation}
where $\epsilon_m$ increases with increasing $m$, $M$
is the total number of eigenvalues for a single realization of the FREO, equal to the number
of trapped modes.

Level spacing statistics can be studied in terms of the random matrix theory
\cite{Stockman}.
Regular dynamics implies that the matrix of the FREO consists
of separate independent blocks. 
Then level sequences contributed from different
blocks are statistically independent, therefore, 
the resulting level spacing distribution obeys the Poisson law
\begin{equation}
  \rho(s)\sim\exp(-s).
\label{Poisson}
\end{equation}
Under conditions of ergodic chaos, 
all normal modes are coupled to each other.
This results in repulsion of neighbouring levels, the phenomenon closely related 
to spectral splittings induced by tunneling \cite{LL3}.
In this case level spacing statistics is described by
the Wigner-Dyson distribution
\begin{equation}
\rho(s)\sim s^{\zeta}\exp\left(-Cs^2\right),
\end{equation}
where constants $\zeta$ and $C$ depend on symmetries
of the FREO.
As the FREO doesn't possess the symmetry 
$r \to -r$, it corresponds to the circular unitary ensemble (CUE)
 with $\zeta=2$ and $C = 4/\pi$ \cite{Kol97}.

\begin{figure}[!ht]
\includegraphics[width=0.45\textwidth]{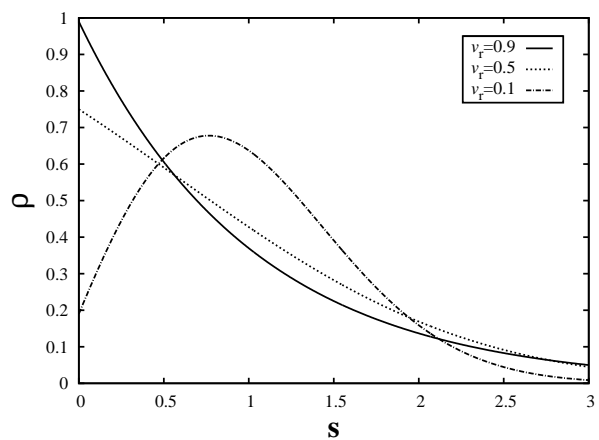}
\caption{
Berry-Robnik distribution with $v_{\mathrm{r}}=0.9$,
$v_{\mathrm{r}}=0.5$ and $v_{\mathrm{r}}=0.1$. 
}%
\label{fig-br_example}
\end{figure}

The most interesting case is
 mixed phase space, with the coexistence
of regular and chaotic domains. 
Then level spacing statistics should be described by some combination
of Poisson and Wigner-Dyson laws.
In the short-wavelength limit one can use
the Berry-Robnik distribution \cite{BR}
\begin{multline}
\rho(s)= \left[
v_{\mathrm{r}}^2\operatorname{erfc}\left(\frac{\sqrt{\pi}}{2}v_{\mathrm{c}}s\right)+\right.\\
\left.+\left(2v_{\mathrm{r}}v_{\mathrm{c}}+\frac{\pi}{2}v_{\mathrm{c}}^3s\right)
\exp\left(-\frac{\pi}{4}v_{\mathrm{c}}^2s^2\right)
\vphantom{\frac{\sqrt{\pi}}{2}}\right]\exp(-v_{\mathrm{r}}s),
\label{berrob}
\end{multline}
where $v_\mathrm{r}$ and $v_\mathrm{c}$ are relative phase space volumes
corresponding to regular and chaotic ray motion, respectively,
 $v_{\mathrm{r}}+v_{\mathrm{c}}=1$.
In (\ref{berrob}), it is assumed that phase space consists of 
only two distinct domains: one regular and one chaotic.

Taking into account the correspondence between rays and modes 
established by the WKB representation (\ref{EBK}), 
one can regard $v_{\mathrm{r}}$ as a 
 fraction of modes whose refraction is regular,
and $v_{\mathrm{c}}$ as a fraction of modes exhibiting wave chaos.
In the limiting cases
 $v_{\mathrm{r}}=1$ and $v_{\mathrm{c}}=1$, 
Berry-Robnik formula (\ref{berrob}) reduces
to the Poisson distribution and the Wigner-Dyson distribution for the orthogonal
ensemble ($\zeta=1$), respectively. 
As is shown in Fig.~\ref{fig-br_example},
Berry-Robnik distribution undergoes smooth transition
from the Poisson to the Wigner-Dyson law as $v_{\mathrm{r}}$ decreases from 1 to 0.
Wigner-Dyson distributions for orthogonal  ($\zeta=1$) and unitary ($\zeta=2$) 
ensembles are close to each other, therefore, one can use (\ref{berrob}) 
as an approximation for the circular unitary ensemble.
Berry-Robnik distribution is based on the assumption of the total 
statistical independence of the matrix blocks corresponding to regular and chaotic dynamics.
This assumption is completely fulfilled only in the semiclassical limit.
As wave corrections grow, independence degrades
due to regular-to-chaotic tunneling \cite{Backer}.
Hence, Berry-Robnik formula cannot work perfectly for low-frequency 
sound propagation.
In this case, a better fit is expected with
the Brody distribution \cite{PR}
\begin{equation}
 \rho(s)=(\beta+1)A_\beta s^\beta\exp(-A_\beta s^{\beta+1}),
\label{brody}
\end{equation}
where $A_\beta=[\Gamma(\frac{\beta+2}{\beta+1})]^{\beta+1}$,
$\Gamma$ is the Euler gamma function. $\beta=0$ corresponds to
the Poisson distribution,
$\beta=1$ yields the Wigner-Dyson distribution with
 $\zeta=1$.
Unfortunately, the Brody distribution is semiempirical
and doesn't have an explicit physical interpretation 
in the intermediate regime $0<\beta<1$.

Functions (\ref{berrob}) and (\ref{brody}) 
should describe level spacing statistics for single realizations of
the FREO. 
Very unfortunately, this is problematic because
one encounters insufficiency of the statistical ensemble.
Indeed, long-range sound propagation is feasible only with
low acoustic frequencies of tens or hundreds Hz. In this frequency range,
number of trapped modes doesn't exceed several hundreds.
To resolve this problem, we consider ensemble-averaged
level spacing distribution
\begin{equation}
\rho(s,\,\tau)=\lim\limits_{N\to\infty}\frac{1}{N}\sum\limits_{n=1}^{N} P_n(s,\,\tau),
\label{aver_ps}
\end{equation}
where $P_n(s,\,\tau)$ is a level spacing distribution
corresponding to  $n$-th realization of FREO.
Fitting the function $\rho(s,\,\tau)$ 
with the Berry-Robnik distribution (\ref{berrob}),
one can estimate number regularly propagating modes for various values
of $\tau$ and, whereby, track the transtion to chaos with increasing $\tau$.
However, it should be noted that formula
 (\ref{aver_ps}) enables accurate estimate of $v_\mathrm{r}$
only  if fluctuations of $v_\mathrm{r}$ are weak, otherwise
one should take into account nonlinearity of
 $\rho$ as a function of $v_\mathrm{r}$ in (\ref{berrob}).

Likewise we can observe transformation of level spacing statistics
 by fitting (\ref{aver_ps}) with the Brody distribution
 (\ref{brody}).
Then the value of $\beta$ corresponding to the best fit should 
increase with increasing $\tau$ from 0 to 1,
reflecting wavefield stochastization.
Such calculations were performed in \cite{Arxiv,UFN} for a waveguide
with a perturbed biexponential sound-speed profile.

\subsection{Eigenfunction statistics}
\label{Eigf}

Wave chaos is also reflected
in eigenfunction statistics of the FREO.
Each eigenfunction can be expressed as a superposition
of normal modes:
\begin{equation}
 \Phi_m(z) = \sum\limits_n c_{mn}\phi_n(z),
\label{eigf}
\end{equation}
where $c_{mn}$ is the $m$-th component of $n$-th eigenvector
of the matrix $\hat G$,
$\phi_n(z)$ is the $n$-th normal mode.

There are many methods for identification of ``chaotic''
eigenfunctions.
In the present work we use only one of them.
Ray chaos can lead to intense energy exchange between normal modes \cite{Viro99,Acoust08,Tomc11},
therefore,
a ``chaotic'' eigenfunction is a compound of many modes. The stronger chaos, the larger
number of contributing modes.
Hence, we can characterize ``chaoticity'' by estimating number of principal components \cite{Varga}
in the expansion (\ref{eigf}).
Number of principal components is calculated as
\begin{equation}
\nu(n) = \left(
\sum\limits_{m=1}^M\lvert c_{mn}\rvert^4
\right)^{-1}.
 \label{npc}
\end{equation}
Number of principal components 
 $\nu$ is equal to 1 in an unperturbed waveguide and grows
as scattering intesifies, tending asymptotically to~$M$.

\section{One-step Poincar\'e map}
\label{Onestep}

In the geometrical acoustics approximation, dynamics of a wavepacket
is represented as motion of some ray bundle.
The bundle remains compact in course of propagation
if ray dynamics is regular.
Under conditions of ray chaos the bundle
rapidly diverges.
Initial conditions for
rays contributing to the bundle can be found by means of
the Husimi function
\begin{multline}
W_\mathrm{h}(p,\,z)=
\biggl|
\frac{1}{\sqrt[4]{2\pi\Delta_z^2}}
\int dz'\Phi_0^*(z')\times \\
\exp\left[ikp(z'-z)-\frac{(z'-z)}{4\Delta_z^2}\right]
\biggr|^2,
\label{Wigner}
\end{multline}
projecting an initial wavepacket $\Phi_0$ onto phase space of ray equations (\ref{sys}). 
In the limit $k_0\to\infty$, 
the transformation (\ref{Wigner}) turns eigenfunctions $\Phi_m$
of the FREO into some phase space sets which are invariant under 
the shift $r=0 \to r=\tau$.
The procedure proposed in \cite{PRE73} allows one
to find out these sets
in a randomly-inhomogeneous waveguide.
This is the one-step Poincar\'e map
(or the specific Poincar\'e map).
The ray analogue of the map  looks as follows:
\begin{equation}
p_{i+1}=p(r=\tau\vert\,p_i,z_i),\quad
z_{i+1}=q(r=\tau\vert\,p_i,z_i),
\label{map}
\end{equation}
where $p(r=\tau\vert\,p_i,z_i)$ and $z(r=\tau\vert\,p_i,z_i)$ are
the solutions of ray equations (\ref{sys}) with
initial conditions $p(r=0)=p_i$, $z(r=0)=z_i$.
Values of $p$ and $z$, calculated at the $i$-th
step of mapping, 
become the initial conditions for the $(i+1)$-th step.
This procedure is equivalent to the usual Poincar\'e map \cite{Zas}
for a range-periodic waveguide with the ray Hamiltonian
\begin{equation}
\bar H=\frac{p^2}{2}+U(z)+\tilde V(z,\,r).
\label{ham-period}
\end{equation}
Here $\tilde V(z,\,r)$ is periodic in  $r$ function  
\begin{equation}
\tilde V(z,\,r'+n\tau)=V(z,\,r'),\quad
0\leqslant r'\leqslant\tau ,
\label{xi-sr}
\end{equation}
where $n$ is an integer. As it follows from (\ref{xi-sr}),
$\tilde V(z,\,r)$ is a sequence of
identical pieces of $V(z,\,r)$, each of them has the length $\tau$.
Thus we replace the original randomly-perturbed Hamiltonian system 
 by an equivalent periodically-perturbed one.
This replacement is valid as long as we 
restrict ourselves by considering dynamics within the range interval
$[0:\tau]$.
One-step Poincar\'e map can be considered as the classical counterpart
of the FREO.

Owing to analogy with the usual Poincar\'e map, 
the main property of the one-step Poincar\'e map
can be formulated as follows:
{\em each point of a continuous closed ray trajectory of the map (\ref{map})
corresponds to a starting point of the solution of (\ref{sys})
which remains stable by Lyapunov till the range $r=\tau$.}
The inverse statement is not, in general, true.
Hence, the one-step Poincar\'e map provides
a sufficient but not necessary criterion of stability.

Map (\ref{map}) was studied in \cite{RayWave,PRE73,JPA,Gan1,Gan2}.
Here we shall give its brief description.
It is reasonable to make canonical transformation of ray variables
from momentum-depth $(p-z)$ to
the action--angle  $(I,\vartheta)$.
This procedure provides more appropriate representation of ray equations.
The angle variable $\vartheta$ canonically conjugated to the action (\ref{action})
is given by
\begin{equation}
 \vartheta=\frac{\partial}{\partial I}
\int\limits_{z_0}^z p\,dz.
\end{equation}
The transformed ray Hamiltonian is written as
\begin{equation}
 \bar H=H_0(I) + \tilde V(I,\vartheta,r).
\end{equation}
Ray equations in terms of the new variables:
\begin{equation}
\begin{gathered}
\frac{dI}{dr}=-\frac{\partial H}{\partial\vartheta}=-\frac{\partial V}{\partial\vartheta},\\
\frac{d\vartheta}{dr}=\frac{\partial H}{\partial I} = \omega(I) + \frac{\partial V}{\partial I},
\end{gathered}
\label{sys-Ith}
\end{equation}
where $\omega=2\pi/D$ is spatial frequency of a ray trajectory in a waveguide.
Perturbation $\tilde V(I,\vartheta)$ can be expanded into a double Fourier series
\begin{equation}
\tilde V = \frac{1}{2}\sum\limits_{k,k'=1}^{\infty}
V_{k,k'}e^{i(k\vartheta - k'\Omega r)} + \text{c.~c.},
 \label{Vseries}
\end{equation}
where $\Omega=2\pi/\tau$. Inserting (\ref{Vseries}) into (\ref{sys-Ith}), we obtain
\begin{equation}
 \begin{aligned}
  \frac{dI}{dr}&=-\frac{i}{2}\sum\limits_{k,k'=1}^{\infty}
  kV_{k,k'}e^{i(k\vartheta-k'\Omega r)} + \text{c.~c.},\\
\frac{d\vartheta}{dr}&=\omega + \frac{1}{2}\sum\limits_{k,k'=1}^{\infty}
\frac{\partial V_{k,k'}}{\partial I}e^{i(k\vartheta-k'\Omega r)} + \text{c.~c.}
 \end{aligned}
\label{sys_Ith1}
\end{equation}
If the condition
\begin{equation}
k'D(I) = k\tau
 \label{rescond}
\end{equation}
is fulfilled, there occurs resonance. The pair of integers
$k'$ and $k$ determines multiplicity of resonance $k:k'$.
Resonances occur at certain values of the action, which correspond
to the so-called resonant torus.
Ray dynamics in a small vicinity of a resonant torus with ray action $I_0$ can be
described using the so-called resonance approximation \cite{Zas}, when one leaves
only resonant terms in the r.h.s. of (\ref{sys_Ith1}).
It should be mentioned that one and the same resonant torus corresponds
to an infinite number of resonances with multiplicities $(jk):(jk')$,
where $j$ is an integer.
However, resonance Fourier amplitudes $V_{k,k'}$ rapidly decrease with increasing
$k$ and $k'$, therefore, only few low-order resonances influence significantly
ray dynamics. 
Consequently, we can take into account only some finite number of dominant resonances.
For further simplification, we make the following procedures:
\begin{enumerate}
\item
As $\tilde V$ is a smooth function of $z$ in the underwater sound channel considered,
the derivative $d\tilde V/dI$ is small compared with $\omega$,
and the sum in the second equation of (\ref{sys_Ith1}) can be dropped out.
\item 
Near resonance, spatial frequency $\omega$ can be expanded as
$\omega=\Omega +\omega_I'\Delta I$, where $\omega_I'=d\omega/dI$.
\end{enumerate}

Then introducing new variables
\begin{equation}
\Delta I = I - I_0,\quad
\psi = k\vartheta - k'\Omega r,
\label{newvar}
\end{equation}
and expressing $V_{k,k'}$ as $\lvert V_{k,k'}\rvert\exp(i\zeta_{k,k'})$, 
we can rewrite (\ref{sys_Ith1}) as
\begin{equation}
\begin{gathered}
\frac{d(\Delta I)}{dr} = \sum\limits_{l=1}^L lk\lvert V_{lk,lk'}\rvert\sin(l\psi + \zeta_{lk,lk'})=-\frac{\partial\tilde H}{\partial\psi},\\
\frac{d\psi}{dr} = \omega_I\Delta I=\frac{\partial\tilde H}{\partial(\Delta I)},
\end{gathered}
 \label{sys-Ith2}
\end{equation}
where $L$ is the number of dominant resonances, and
\begin{equation}
\tilde H=\omega_I'\frac{(\Delta I)^2}{2}+\sum\limits_{l=1}^L
 lk\lvert V_{lk,lk'}\rvert\cos(l\psi + \zeta_{lk,lk'}).
 \label{universal}
\end{equation}
If $L=1$, (\ref{universal}) turns into the universal Hamiltonian of nonlinear resonance \cite{RayWave,Zas},
and a phase space portrait 
of Eqs.~(\ref{sys-Ith2}) contains the domain of finite  motion enclosed
by the separatrix and corresponding to trapping into resonance.
Then maximal value $\Delta I$ on the separatrix is determined as
\begin{equation}
 \Delta I_\text{max} = 4\sqrt{\frac{k\lvert V_{k,k'}\rvert}{\omega_I'}}
\label{width}
\end{equation}
The terms with $l>1$ may deform the pendulum-like phase space portrait and, moreover,
result in the presence of additional separatrices inside the domain
of finite motion. The latter phenomenon can occur 
when the perturbation oscillates with depth \cite{PRE73}.

Transition to global chaos in the one-step Poincar\'e map happens when 
neighbouring dominant resonances overlap.
The criterion of overlapping is the well-known Chirikov criterion
\begin{equation}
\frac{\Delta I_\text{max}}{\delta I} \geqslant 1.
 \label{Chirikov}
\end{equation}
Here $\delta I$ is the distance between neighbouring dominant resonances in the action space.
Its variability with $\tau$ for $\tau>D$ is described by equation
\begin{equation}
 \delta I = \frac{2\pi}{\omega_I'\tau},
\end{equation}
that is, increasing of $\tau$ enhances resonance overlapping.
Differences in phase space patterns corresponding to different realizations of the perturbation
are associated with phase and amplitude fluctuations
of Fourier-amplitudes $V_{k,k'}$. 
However, the contribution of these fluctuations is limited, 
therefore,
the ratio of the phase space volumes corresponding to regular and chaotic motion
is mainly controlled by $\tau$ and weakly varies from one realization to another
(see below for an illustration).
This property
allows one to consider the one-step Poincar\'e map as an useful tool
for studying randomly-driven dynamical systems of various physical origins.

Resonance approximation presented above fails under violation of the nondegeneracy
condition $\omega_I'\ne 0$. In this case one should use more sophisticated approaches, 
one of them is presented in \cite{Rypina}. 
We shall address this issue in the end of this Section.

\begin{figure*}[!htb]
\begin{center}
\includegraphics[width=0.8\textwidth]{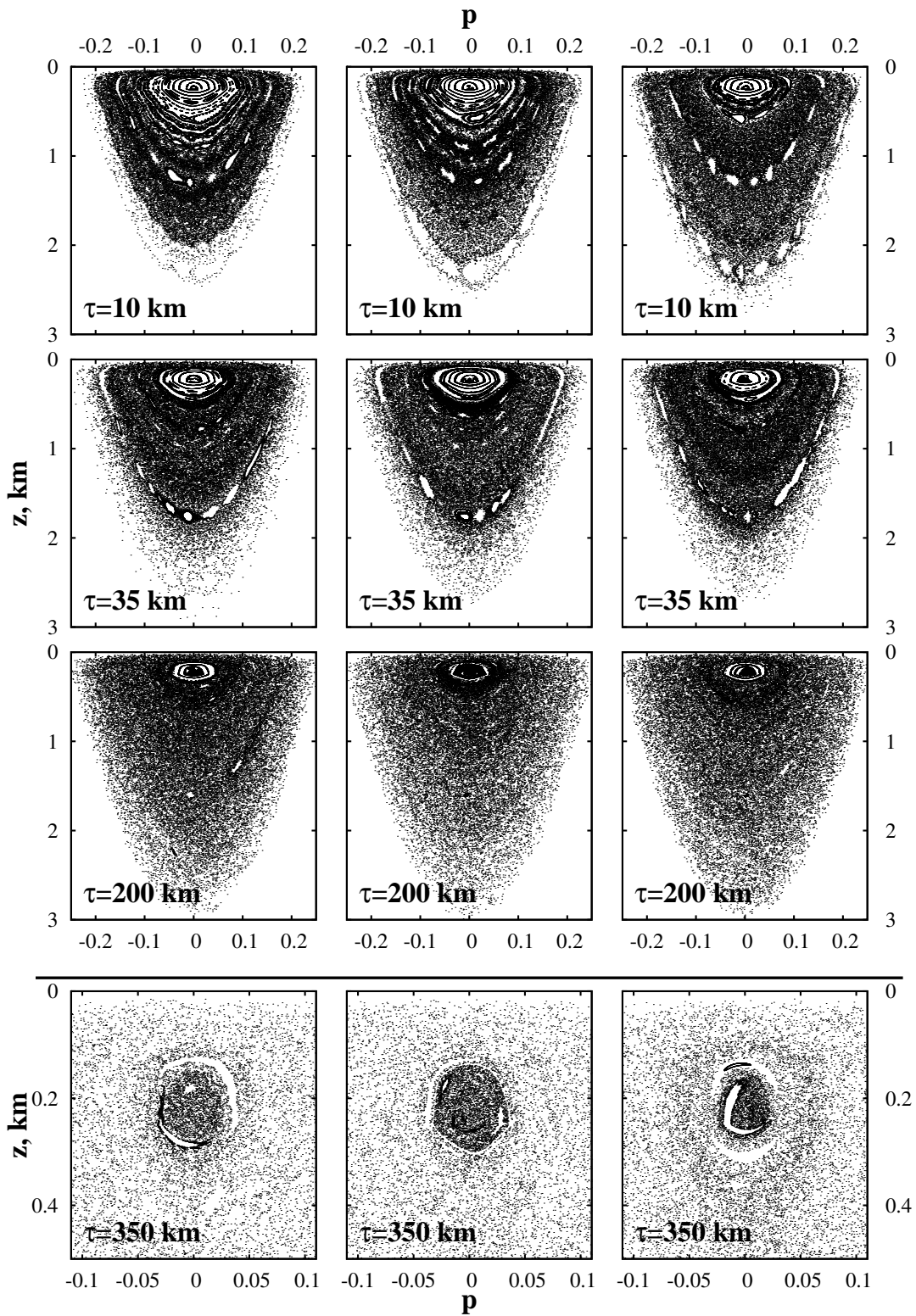}
\caption{Ray phase space portraits constructed via the one-step Poincar\'e map (\ref{map}).
Each column corresponds a single realization of the sound-speed perturbation.
Value of $\tau$ is indicated in the left lower corner.
}%
\label{fig-poinc}
\end{center}
\end{figure*}

Fig.~\ref{fig-poinc} illustrates phase space portraits calculated
using the one-step Poincar\'e map 
with three different realizations of the sound-speed perturbation.
The jump of the derivative
 $dU/dz$ at $z=z_0$ leads to fast growth of numerical error in ray calculations.
Therefore, we replaced the expression (\ref{uz}) for $U(z)$ by 
the smoothed function
\begin{equation}
 U(z) = U_1(z) + \frac{1}{2}\left[1+\tanh\frac{z-z_0}{\Delta}\right]U_2(z),
\end{equation}
where $\Delta=1$~m.
Each of the phase space portraits represents a mixed phase space
structure consisted of regular and chaotic domains.
Phase portraits with the same $\tau$
mainly differ only in angular locations of regular islands, 
whereas their overall structure is very similar.
The main regular domain is placed near the point $z=z_0$, $p=0$
and corresponds to flat rays intersecting the horizontal plane
with the smallest angles.
This circumstance deserves especial attention because stability of flat rays 
is not typical for sound propagation in the deep ocean.
Numerous experiments on long-range sound propagation in the North-Eastern
Pacific Ocean (see, for instance, \cite{Kaneohe,SLICE89,Worc-AET,AST,LOAPEX})
indicate on strong irregularity of flat near-axial rays, associated with ray chaos
\cite{Simmen,AET}.
The ``deterministic'' mechanism of near-axial chaos is ray scattering
on vertical resonances caused
by small-scale depth oscillations of the sound-speed perturbation \cite{Hege,PRE76,Acoust08}.
These oscillations are contributed from high-number modes of an internal-wave field.
In the Sea of Japan, the effect of the high-number modes is weak, therefore,
the sound-speed perturbation can be fairly described by equation
 (\ref{factoriz}), where depth dependence is given by
a smooth function $Y_1(z)$.
Weakness of high-number internal-wave modes is a peculiar feature 
of the Sea of Japan.
It is caused by the specific form of the buoyancy
frequency profile \cite{IEEE}. In particular, the waveguide for internal waves,
determined by the buoyancy frequency profile, is too narrow. 
Consequently, it doesn't efficiently focus  high-number modes
of low-frequency internal waves.
As low-frequency internal waves give the dominant contribution
into a total internal-wave field, 
depth oscillations of the perturbation are suppressed.

Resonance overlapping is enhancing as $\tau$ grows,
and stable islands eventually submerge into the chaotic sea.
However, there is a small region of stability that survives for distances of hundreds kilometers,
transforming into a chain of islands around the region near $z=z_0$, $p=0$.
This chain corresponds to the smooth minimum of the function $D(p_0)$ depicted
in Fig.~\ref{fig-dp}.
In the absence of inhomogeneity, this minimum gives rise to the
so-called shearless torus in phase space,
producing a weakly-divergent beam. 
It is recognized that shearless tori can possess extraordinary 
persistence to chaos \cite{Rypina,Nontwist,Bud79,Bud81,JETP10}, therefore,
a weakly-divergent beam can survive in the presence of a sound-speed perturbation.
Hence, formation of a weakly-divergent beam can be considered as a possible
mechanism responsible for unusual stability of near-axial rays, observed in experiments \cite{Bez09,IEEE}.

\section{Spectral statistics of FREO for the underwater sound channel in the Sea of Japan}
\label{Japan}

\subsection{General remarks}
\label{Gennotes}

\begin{figure}[!tb]
\begin{center}
\includegraphics[width=0.48\textwidth]{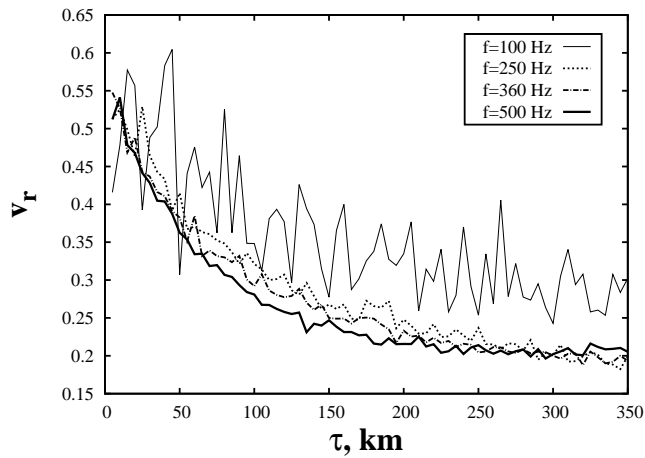}
\caption{
Fraction of the phase space volume corresponding to regular motion
vs distance $\tau$ for various frequencies.
}%
\label{fig-berry}
\end{center}
\end{figure}

This section is devoted to numerical modeling of the FREO
in the Sea of Japan.
Each realization of the FREO was represented as a matrix
in the basis of normal modes
being solutions of the Sturm-Liouville problem (\ref{StL})
with the boundary conditions (\ref{BCs}).
Only purely-water modes which propagate without reaching the bottom
were taken into account. They were selected using the criterion
$E_m<U(z=h)$, where $E_m$ is the $m$-th eigenvalue of the Sturm-Liouville problem
 (\ref{StL}).
This criterion ensues from the WKB approximation for normal modes \cite{Viro99}.
Number of trapped modes $M$ depends on sound frequency.
It is equal to 72 for $f=100$~Hz, 179 for $f=250$~Hz,  
259 for $f=360$~Hz, and 361 for $f=500$~Hz.
Statistical ensembles of the FREOs, corresponding to the frequencies of 250, 360 and 500~Hz, were calculated with
100 realizations of the perturbation. The ensemble corresponding to 100 Hz was calculated with
500 realizations.
For each realization, we constructed a family of the FREOs 
$\hat G(\tau)$, where $\tau=5,10,15,\dots,350$~km.

\begin{figure}[!htb]
\includegraphics[width=0.48\textwidth]{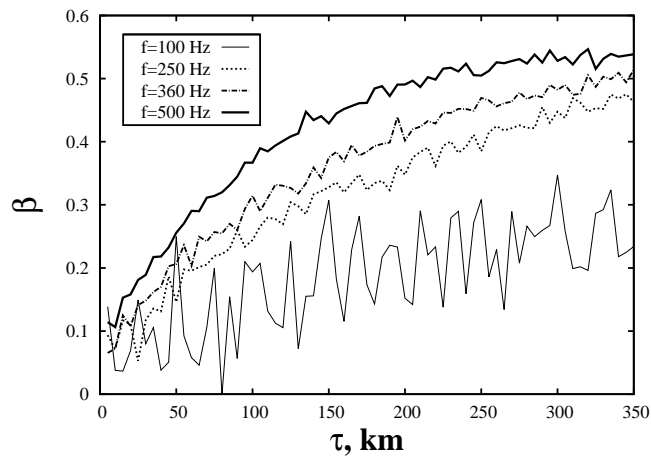}
\caption{The Brody parameter $\beta$ vs distance $\tau$.
}%
\label{fig-brody}
\end{figure}

\subsection{Eigenvalue statistics}
\label{Eigv-j}


We calculated the ensemble-averaged
level spacing distribution $\rho(s,\tau)$ 
using the formula (\ref{aver_ps})
and fitted it, for each value of 
$\tau$, by means of the Berry-Robnik  (\ref{berrob})
and Brody (\ref{brody}) distributions.
Thus, we obtained dependencies of the regular phase space volume
  $v_{\mathrm{r}}$ and Brody parameter $\beta$
on distance $\tau$.
As is shown in Fig.~\ref{fig-berry},
$v_{\mathrm{r}}$ rapidly decreases in the first 100--150~km.
Then, it becomes almost constant. 
This may indicate on the influence
of the long-living stable islands in the vicinity of the weakly-divergent beam.
Notably, the curves corresponding to 250, 360 and 500~Hz
are very close to each other, whereas
the curve corresponding to
100 Hz lies above them and undergoes strong fluctuations
which persist even with increasing number of realizations.
It should be mentioned that 
the above estimates of $v_{\mathrm{r}}$ have limited accuracy
because the assumptions underlying the formula (\ref{berrob})
are satisfied only approximately. Therefore, the results
obtained using the Berry-Robnik distribution are rather 
qualitative than quantitative, especially for low sound frequencies.

\begin{figure}[!htb]
\includegraphics[width=0.48\textwidth]{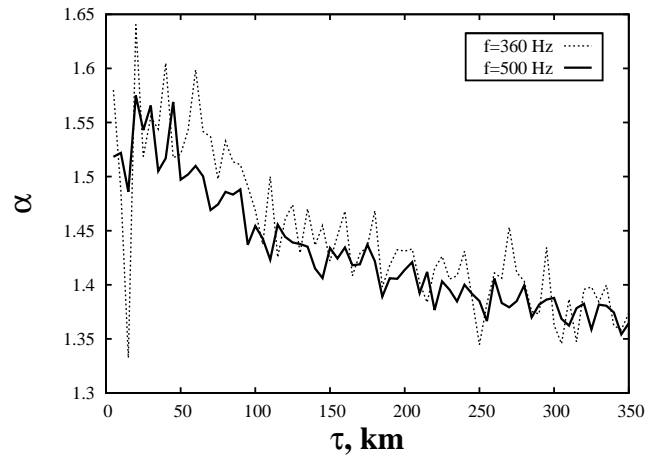}
\caption{
Relano parameter $\alpha$ vs distance $\tau$.
}%
\label{fig-relano}
\end{figure}

Approximation of level spacing statistics by means of the Brody distribution
leads to qualitatively similar results. They are presented in Fig.~\ref{fig-brody}.
For frequencies 250, 360 and 500~Hz,
the Brody parameter $\beta$ gradually grows from 0 to~1,
reflecting transformation from Poissonian to Wigner-like statistics.
In the case of $f=100$~Hz the growth is remarkably slower and affected by strong fluctuations.
The fluctuations of $v_{\mathrm{r}}$ and $\beta$
may be induced by regular-to-chaotic tunneling and dynamical localization,
the phenomena whose influence on spectral statistics is still not well understood.

In addition, we used the method of spectral analysis, developed
by A.~Relano with coworkers in \cite{Relano02}.
In this method, one firstly constructs a series
\begin{equation}
 \delta_n=\sum\limits_{i=1}^n(s_i-\left<s\right>),
\label{sequence}
\end{equation}
where $n=1,2,\dots,N-1$, $N$ is the total number of eigenvalues.
Then, making discrete Fourier transform
\begin{equation}
 \bar\delta_k=\frac{1}{\sqrt{N}}\sum\limits_n \delta_n\exp\left(
\frac{2\pi ikn}{N}
\right),
\label{relano-f}
\end{equation}
one finds the power spectrum
\begin{equation}
 S(k)=\lvert\bar\delta_k\rvert^2.
\label{S_k}
\end{equation}
Generally, ensemble-averaged spectrum obeys a power low
\begin{equation}
\left<S(k)\right>\sim k^{-\alpha}.
\label{alfa}
\end{equation}
Relano with coworkers found that regular dynamics corresponds to 
 $\alpha=2$, and global chaos results in $\alpha=1$.
In the mixed regime,  $\alpha$ takes on an intermediate value between
1 and 2 \cite{Relano08}.
Fig.~\ref{fig-relano} demonstrates that
 $\alpha$ decreases with increasing $\tau$ for the frequencies of
 360 and 500~Hz, indicating gradual transition to chaos.
However,  $\alpha$ varies relatively slowly and remains 
near the middle value 1.5 for all distances considered, despite of the marked changes 
in the classical phase space portrait (see Fig.~\ref{fig-poinc}).
Analogous dependencies for the frequencies of 100 and 250 Hz exhibit strong fluctuations
and, therefore, are not presented in the figure.
This implies that the method developed in \cite{Relano02,Relano08} provides
good agreement only for relatively short wavelengths.

 \subsection{Eigenfunction analysis}
\label{Eigf-j}

Analysis of eigenfunctions possesses some advantages
as compared with analysis of eigenvalues.
The main advantage is the possibility
to associate each eigenfunction with some set of normal modes and, whereby,
associate it with a certain geometry of propagation.
We can properly classify eigenfunctions,
taking into account the interplay with normal modes.
Such classification can be used
for finding wavepacket configurations whose dynamics 
is expected to be less or more regular.
Proper classification can be obtained using the parameter $\mu$
\cite{Viro05}. It is defined as
\begin{equation}
\mu=\sum\limits_{m=1}^M\lvert c_m\rvert^2m.
 \label{mu}
\end{equation}
In an unperturbed waveguide,
only one normal mode contributes to each eigenfunction, and
$\mu$ coincides with the number of this mode.
Taking into account the quantization rule (\ref{EBK}),
we obtain the formula
\begin{equation}
\left<I\right>=\frac{\mu}{k_0}+\frac{1}{2k_0}
\label{mean_I}
\end{equation}
that gives mean action corresponding to an eigenfunction.
According to (\ref{mean_I}),
the parameter $\mu$ determines phase space location
of the eigenfunction and can serve as its identificator.
%
\begin{figure*}[!tb]
\begin{center}
\includegraphics[width=0.8\textwidth]{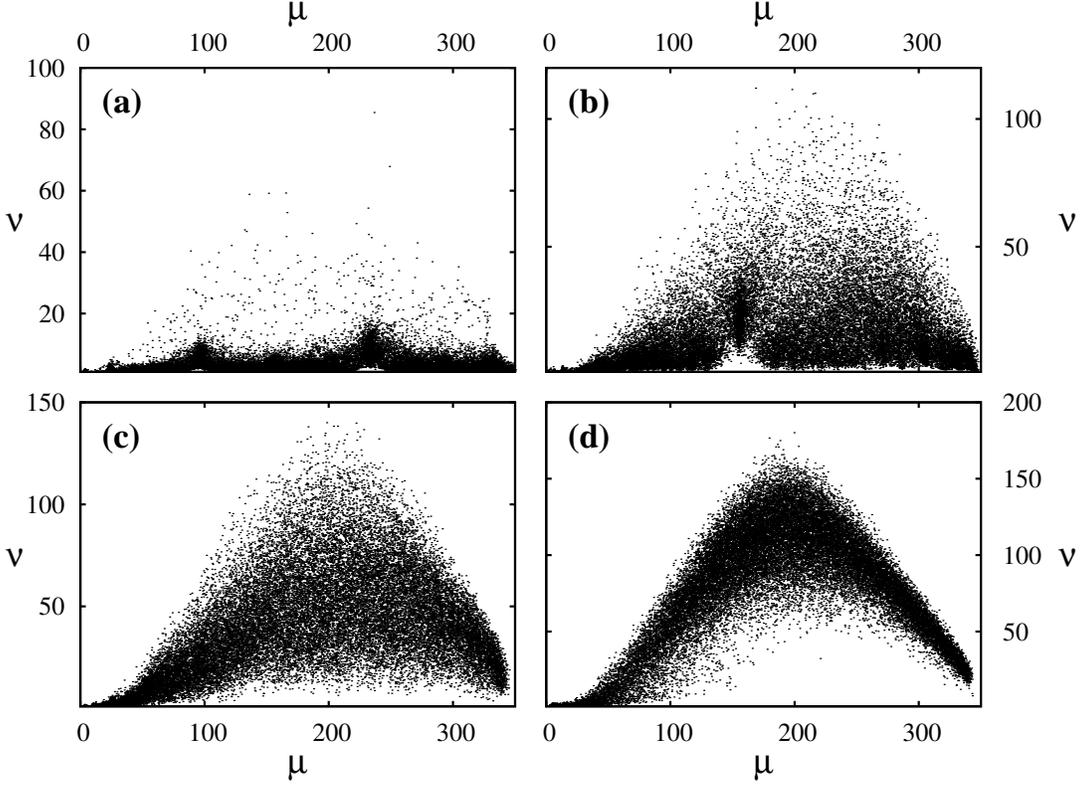}
\caption{Distribution of eigenfunctions in the $\mu$--$\nu$ plane, 
where the parameter $\mu$ is given by (\ref{mu}), and $\nu$ is a number 
of principal components (\ref{npc}).
Distance values:
(a) $\tau=10$~km, (b) $\tau=35$~km, (c) $\tau=100$~km, (d) $\tau=350$~km.
The sound frequency is 500~Hz.
}%
\label{fig-npc500}
\end{center}
\end{figure*}
Figures \ref{fig-npc500} and \ref{fig-npc100}
illustrate eigenfunction distributions in the $\mu$--$\nu$ plane,
where $\nu$ is number of principal components (\ref{npc}),
for the frequencies of 500 and 100 Hz, respectively.

Let us firstly consider Fig.~\ref{fig-npc500} corresponding to the frequency of 500 Hz.
Mode coupling is relatively weak for small values of $\tau$, therefore,
the distribution  is mainly concentrated
near  $\nu=1$.
We want to focus attention on
the dense spots elongated in the $\nu$-direction.
They correspond to eigenfunctions reflecting mode-medium resonance \cite{Viro99}
being the wave counterpart of ray-medium resonance (\ref{rescond}).
Indeed, 
an isolated ray-medium resonance produces oscillations of ray action
inside the interval $I_0-\Delta I_\text{max}\leqslant I \leqslant I_0+\Delta I_\text{max}$,
where $I_0$ is a resonance action, $\Delta I_\text{max}$ is 
resonance width in the action space.
According to the principle of ray-mode duality,
these oscillations of action correspond to coherent transitions between the normal modes
whose numbers satisfy the inequality
\begin{equation}
m_0-\Delta m \leqslant m \leqslant m_0 + \Delta m,
\end{equation}
where $m_0=k_0I_0 + 1/2$, $\Delta m=k_0\Delta I_\text{max}$.
Resonance-induced modal transtions give rise 
to eigenfunctions with $\mu\simeq m_0$ and $\nu$ varying from 1 to $\Delta m$.
As long as resonance values of the action, being determined by $\tau$, 
are the same for all realizations of the perturbation, 
these eigenfunctions form vertically-elongated concentrations of points
in the $\mu$--$\nu$ plane.
Location of mode-medium resonances along the $\mu$-axis can be found
using the formula
\begin{equation}
k'D(I=\left<I\right>) = k\tau ,
 \label{mode-medium}
\end{equation}
where $\left<I\right>$ is linked to $\mu$ by (\ref{mean_I}).
Concentrations induced by mode-medium resonance disappear 
with increasing $\tau$ due to overlapping of mode-medium resonances
and delocalization \cite{BerKol}. Delocalization
leads to abrupt growth of number of principal components.
It eventually subjects all eigenfunctions in the interval between
 $\mu\simeq100$ and $\mu\simeq300$,
resulting in the ``boomerang'' pattern 
in the $\mu$--$\nu$ plane, as illustrated in Fig.~\ref{fig-npc500}(d).
Left and right ends of the ``boomerang'' are formed by 
weakly scattered eigenfunctions. 
The left end corresponds to the almost horizontal near-axial propagation,
that is, its regularity can be associated with
long-living stable islands in the one-step Poincar\'e map. 

%
\begin{figure*}[!tb]
\begin{center}
\includegraphics[width=0.8\textwidth]{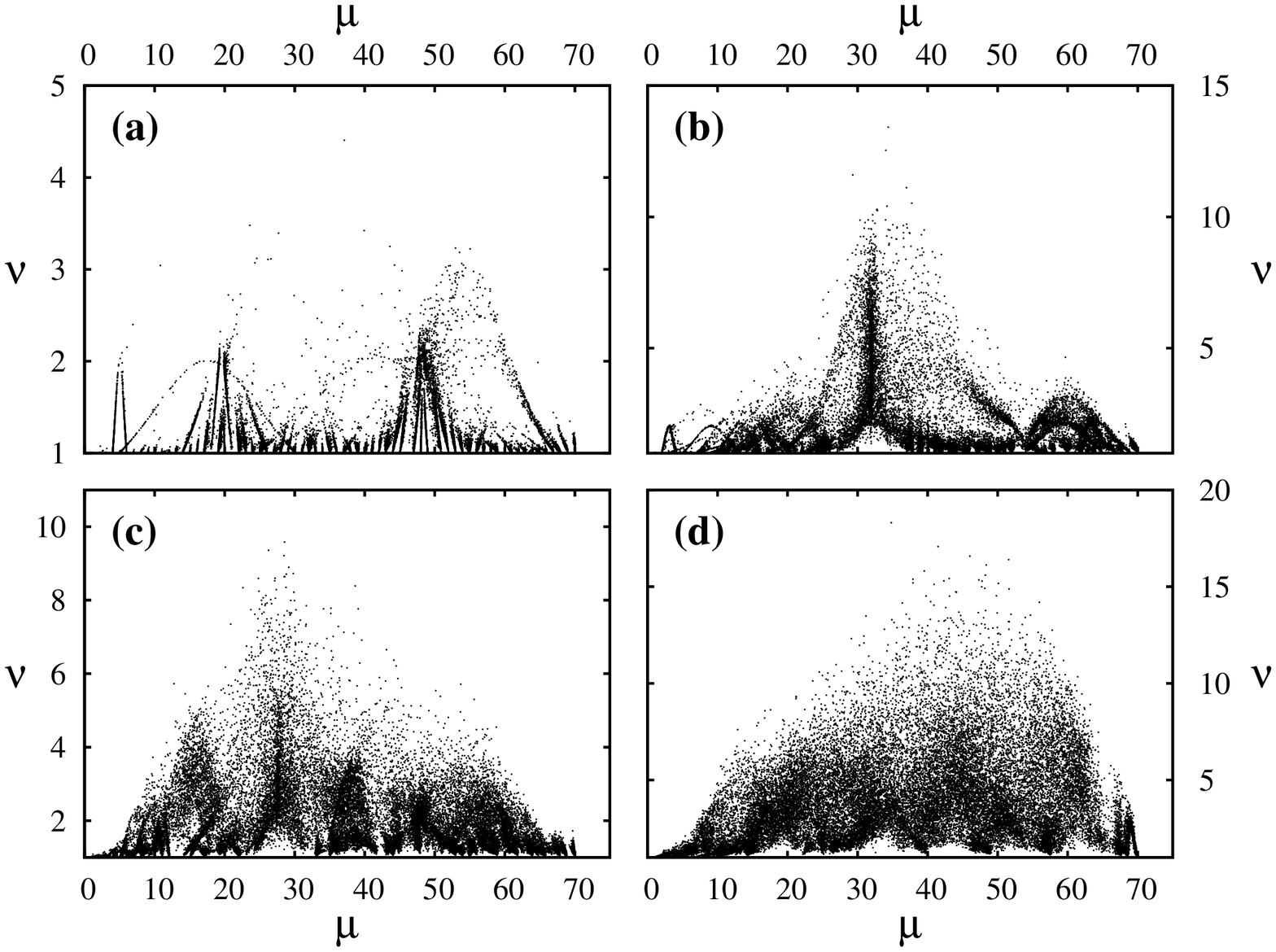}
\caption{The same as in Fig.~\ref{fig-npc500},
but for the frequency of 100 Hz.
(a) $\tau=10$~km, (b) $\tau=35$~km, (c) $\tau=100$~km, (d) $\tau=350$~km.
}%
\label{fig-npc100}
\end{center}
\end{figure*}

Eigenfunction distribution in the $\mu$--$\nu$ plane for the frequency
of 100~Hz possesses a more complicated structure.
It is exceptionally regular for $\tau=10$~km and $\tau=35$~km,
as shown in Figs.~\ref{fig-npc100}(a) and \ref{fig-npc100}(b).
Mode-medium resonances reveal themselves as ``stalagmites''.
Each stalagmite is drawn by a family of distinct weakly biased lines.
Contours of the most pronounced stalagmite
for  $\tau=35$~km are somewhat disordered and smeared.
Some traces of stalagmite-like patterns survive
even for distances of hundreds kilometers, despite of 
global overlapping of ray-medium resonances.
Persistence of stalagmites for large $\tau$ indicates
on the presence of eigenstates localized near periodic
orbits of the one-step Poincar\'e map.
As long as these eigenstates correspond to nonspeading
wavepackets, we can associate such localization with
suppresion of ray chaos and
recovery of regular refraction in the vicinities of the periodic orbits.
An analogous phenomenon had been reported in \cite{PRE76}
for range-periodic waveguides.

Besides of stalagmites, Figs.~\ref{fig-npc100}(a)
and \ref{fig-npc100}(b) illustrate the pattens in the form of ``bridges''.
For instance, a pronounced ``bridge'' in the left part of 
Fig.~\ref{fig-npc100}(a) 
connects the points $\mu=5$, $\nu=1$ and $\mu=30$, $\nu=1$.
The eigenfunctions producing the ``bridges''
are consisted of normal modes satisfying
the condition
\begin{equation}
k_0(E_m-E_n)=\frac{2\pi l}{\tau},\quad m>n.
 \label{resc-q}
\end{equation}
The aforementioned ``bridge'' in the Fig.~\ref{fig-npc100}(a)
satisfies (\ref{resc-q}) with $m=30$, $n=5$, and $l=9$.
Condition (\ref{resc-q}) is equivalent to quantum resonance
between two energy levels.
In the ray limit
\begin{equation}
 k_0(E_m-E_n)\to (m-n)\frac{dE}{dI}\equiv \frac{2\pi(m-n)}{D},
\end{equation}
and condition (\ref{resc-q}) reduces to (\ref{rescond}).
Each realization of the FREO 
can yield eigenfunctions manifesting resonance (\ref{resc-q}). 
If resonance (\ref{resc-q}) corresponding to some numbers ($l$, $m$, $n$)
is localized, i.~e. the modes $m$ and $n$ are not affected by other resonances,
then the respective eigenfunction is a superposition
of normal modes $m$ and $n$,
\begin{equation}
 \Phi_\text{res}(z)\simeq c_m\phi_m + c_n\phi_n,\quad 
\lvert c_m\rvert^2 + \lvert c_n\rvert^2 \simeq 1.
\label{resfunc}
\end{equation}
As this takes place,
the ratio of amplitudes
$\lvert c_m\rvert /\lvert c_n\rvert $ is determined by the phase of the resonance harmonics of the perturbation.
For the perturbation (\ref{factoriz}),
amplitude of the $l$-th resonance harmonics reads
\begin{equation}
 B_l=\frac{1}{2\pi}\int\limits_{0}^{2\pi/\tau} b_1(r)\exp\left(-i\frac{2\pi lr}{\tau}\right)dr.
\end{equation}
Phase of the resonance harmonics is a random quantity with uniform probability density
in the range $[0:2\pi]$. 
Each value of the phase uniquely determines the values of $\mu$ and $\nu$
via the formulae
\begin{equation}
 \mu = \lvert c_m\rvert^2m + \lvert c_n\rvert^2n,
\end{equation}
\begin{equation}
 \nu = \left(\lvert c_m\rvert^4 + \lvert c_n\rvert^4\right)^{-1}.
\end{equation}
It turnes out that quantities $\mu$ and $\nu$ are correlated
for each eigenfunction corresponding to localized resonance (\ref{resc-q}). 
This results in formation of ``bridges'' in the $\mu$--$\nu$ plane.
Localization becomes violated with increasing $\tau$, therefore,
correlation between $\nu$ and $\mu$ ceases, and 
ordered ``bridges'' transform into disordered clouds of points.

\begin{figure}[!tb]
\begin{center}
\includegraphics[width=0.48\textwidth]{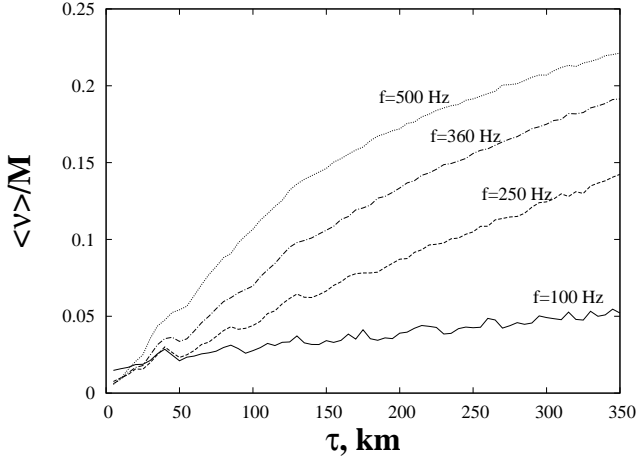}
\caption{Mean number of principal components as function
of distance.
}%
\label{fig-mean_npc}
\end{center}
\end{figure}

The above analysis shows that there are qualitative differences 
in sound scattering for different frequencies.
To estimate these differences quantitatively, we need 
a suitable parameter characterizing scattering strength.
For example, chaos-induced phase space delocalization
can be measured by the ensemble-averaged number of principal
components, divided by the total number of trapped modes.
As it follows from Fig.~\ref{fig-mean_npc}, the rate of delocalization
increases with increasing sound frequency.
This implies that chaotic diffusion associated with ray chaos
degrades with increasing sound wavelength.
It can be thought of as a manifestation of dynamical localization \cite{Stockman},
an analogue of Anderson localization, when destructive interference
supresses wavepacket spreading.

\begin{figure}[!tb]
\begin{center}
\includegraphics[width=0.48\textwidth]{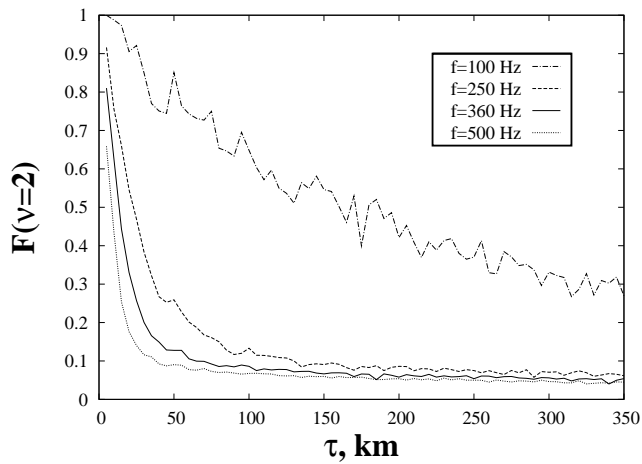}
\caption{Fraction of strongly-localized eigenfunctions
as function of distance.
The criterion of strong localization is the inequality
$\nu\leqslant 2$.
}%
\label{fig-F_2}
\end{center}
\end{figure}

In practice, it is useful to know 
fraction of the eigenfunction ensemble corresponding to regular propagation.
This quantity can be regarded as an analogue of the parameter $v_{\mathrm{r}}$
in the Berry-Robnik distribution.
It can be estimated by means of the cumulative distribution function
\begin{equation}
F(\nu)=\int\limits_{1}^{\nu}\rho(\nu')\,d\nu',
\label{cumul} 
\end{equation}
where $\rho(\nu')$ is the probability density function of $\nu$.
We can conditionally distinguish two regimes of localization:
strong localization and moderate localization.
Strong localization implies that eigenfunction of FREO is close
to one of normal modes of the unperturbed waveguide.
To select strongly-localized eigenfunctions, we can use the inequality
$\nu\leqslant 2$.
Dependence of $F(2)$ on the horizontal distance $\tau$ is depicted
in Fig.~\ref{fig-F_2}.
Evidently, fraction of strongly-localized eigenfunctions
is much larger for  $f=100$~Hz than for higher frequencies.
The curves corresponding to the higher frequencies are close
to each other. They drop down in the range of small $\tau$,
 then decreasing of $F(2)$ becomes very slow.
Slow decreasing can be linked to the presence of the long-living
islands of stability in the neibourhood of the weakly-divergent beam.

\begin{figure}[!tb]
\begin{center}
\includegraphics[width=0.48\textwidth]{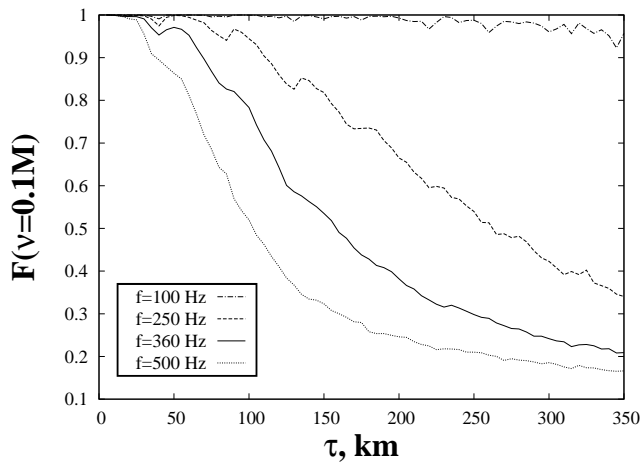}
\caption{Fraction of moderately localized eigenfunctions
vs distance. 
The criterion of moderate localization is the inequality
$\nu\leqslant 0.1M$, where $M$ is the number of trapped modes.}%
\label{fig-F_d}
\end{center}
\end{figure}

In the regime of moderate localization,
mode coupling can be sufficiently strong, but an eigenfunction
occupies relatively small phase space volume, that is,
number of principal
components is limited.
We use the inequality $\nu\leqslant 0.1M$
as the criterion of moderate localization.
As is demonstrated in Fig.~\ref{fig-F_d}, 
almost all eigenfunctions corresponding to the frequency of 100~Hz 
are moderately-localized.
This is not the case of higher frequencies, when fraction
of moderately-localized eigenfunctions significantly decreases with $\tau$.
We can conclude that decreasing of sound frequency
can result in remarkable suppression of wavepacket diffusion
in phase space.

\section{Conclusion}

Spectral analysis of the finite-range evolution operator
prompts a way to explore wave dynamics in a randomly-inhomogeneous
waveguide by means of the quasideterministic approach,
involving resonances, periodic orbits, phase space portraits, e.t.c.
This approach was originally proposed in \cite{UFN,Arxiv}.
The present work is devoted to its further development.
We demonstrate various methods of spectral analysis.
For instance, 
fitting of level spacing  statistics by means of
the Berry-Robnik distribution yields
approximate estimate for
fraction of regularly propagating normal modes of a waveguide.
Also we use the method developed by A.~Relano with coworkers,
and consider distribution of eigenfunctions 
in the $\mu$--$\nu$ space.
In our opinion, the latter approach is the most robust,
albeit its analytical description is lacking.
The most important advantage
of the eigenfunction analysis is the
possibility to study separately scattering of
different modes of a waveguide and 
find out the modes corresponding to regular propagation.
A detailed view of the eigenfunction distribution
in the $\mu$--$\nu$ space
suggests that the mechanism of the chaos onset with increasing distance
can be associated with 
overlapping and delocalization of mode-medium resonances.
The approach based on the statistical analysis of level spacings
by means of the Berry-Robnik formula gives
qualitative description of the transition but doesn't
ensure quantitative agreement.
The Relano method and the method based on the Brody distribution
also give the qualitative description but cannot 
make any quantitative estimates due to their semiempirical
nature.
 
We consider the underwater sound channel in the Sea of Japan 
as an example. Our analysis shows that
almost horizontal near-axial sound propagation
 preserves regularity over distances of hundreds
kilometers. There are two factors responsible for near-axial stability.
The first factor is the peculiar hydrological structure in the region
considered, resulting in the absence of vertical oscillations of the sound-speed
perturbation. 
This circumstance leads to a qualitatively different scenario
of ray chaos, as compared with 
the well-known acoustic experiments in the North-Eastern Pacific Ocean.
The second factor is the formation of the weakly-divergent beam
supported by long-living stable islands
in classical phase space. Weakly-divergent beam occurs
in the vicinity of the shearless torus. Non-dispersive motion of quantum wavepackets
near shearless tori was earlier observed  in \cite{Kudo}.
Also, we should emphasize that sound propagation with the frequency of 100 Hz reveals
exceptionally high degree of stability, 
indicating on different physics of scattering for low sound frequencies.
This phenomenon can be associated with strong dynamical localization.

We suppose that applicability of the approach presented in this paper
is not limited to the problems of wave propagation in random media.
An analogue of the FREO can be readily used for studying of noise-induced
quantum transport and related phenomena. We suggest that combination
of deterministic and statistical approaches should provide an insightful view
on details of dynamics, especially in the presence of intermittency or
synchronization.

This work was supported by the Russian Foundation of Basic Research
(projects 09-05-98608 and 09-02-01258-a), 
the Federal Program ``World Ocean'' and the ``Dynasty'' Foundation.
Authors are grateful to A.R.~Kolovsky,
S.~Tomsovic, O.A.~Godin, K.V.~Koshel and V.V.~Novotryasov for 
helpful comments concerning the subject of the research.

\bibliographystyle{apsrev4-1}
\bibliography{biblio}

\begin{thebibliography}{61}%
\makeatletter
\providecommand \@ifxundefined [1]{%
 \@ifx{#1\undefined}
}%
\providecommand \@ifnum [1]{%
 \ifnum #1\expandafter \@firstoftwo
 \else \expandafter \@secondoftwo
 \fi
}%
\providecommand \@ifx [1]{%
 \ifx #1\expandafter \@firstoftwo
 \else \expandafter \@secondoftwo
 \fi
}%
\providecommand \natexlab [1]{#1}%
\providecommand \enquote  [1]{``#1''}%
\providecommand \bibnamefont  [1]{#1}%
\providecommand \bibfnamefont [1]{#1}%
\providecommand \citenamefont [1]{#1}%
\providecommand \href@noop [0]{\@secondoftwo}%
\providecommand \href [0]{\begingroup \@sanitize@url \@href}%
\providecommand \@href[1]{\@@startlink{#1}\@@href}%
\providecommand \@@href[1]{\endgroup#1\@@endlink}%
\providecommand \@sanitize@url [0]{\catcode `\\12\catcode `\$12\catcode
  `\&12\catcode `\#12\catcode `\^12\catcode `\_12\catcode `\%12\relax}%
\providecommand \@@startlink[1]{}%
\providecommand \@@endlink[0]{}%
\providecommand \url  [0]{\begingroup\@sanitize@url \@url }%
\providecommand \@url [1]{\endgroup\@href {#1}{\urlprefix }}%
\providecommand \urlprefix  [0]{URL }%
\providecommand \Eprint [0]{\href }%
\providecommand \doibase [0]{http://dx.doi.org/}%
\providecommand \selectlanguage [0]{\@gobble}%
\providecommand \bibinfo  [0]{\@secondoftwo}%
\providecommand \bibfield  [0]{\@secondoftwo}%
\providecommand \translation [1]{[#1]}%
\providecommand \BibitemOpen [0]{}%
\providecommand \bibitemStop [0]{}%
\providecommand \bibitemNoStop [0]{.\EOS\space}%
\providecommand \EOS [0]{\spacefactor3000\relax}%
\providecommand \BibitemShut  [1]{\csname bibitem#1\endcsname}%
\let\auto@bib@innerbib\@empty
\bibitem [{\citenamefont {Shockley}\ \emph {et~al.}(1982)\citenamefont
  {Shockley}, \citenamefont {Northrop}, \citenamefont {Hansen},\ and\
  \citenamefont {Hartdegen}}]{Shockley}%
  \BibitemOpen
  \bibfield  {author} {\bibinfo {author} {\bibfnamefont {R.~C.}\ \bibnamefont
  {Shockley}}, \bibinfo {author} {\bibfnamefont {J.}~\bibnamefont {Northrop}},
  \bibinfo {author} {\bibfnamefont {P.~G.}\ \bibnamefont {Hansen}}, \ and\
  \bibinfo {author} {\bibfnamefont {C.}~\bibnamefont {Hartdegen}},\ }\href
  {\doibase 10.1121/1.387250} {\bibfield  {journal} {\bibinfo  {journal} {J.
  Acoust. Soc. Am.}\ }\textbf {\bibinfo {volume} {71}},\ \bibinfo {pages} {51}
  (\bibinfo {year} {1982})}\BibitemShut {NoStop}%
\bibitem [{\citenamefont {Munk}\ \emph {et~al.}(1988)\citenamefont {Munk},
  \citenamefont {O'Reilly},\ and\ \citenamefont {Reid}}]{MunkRR}%
  \BibitemOpen
  \bibfield  {author} {\bibinfo {author} {\bibfnamefont {W.~H.}\ \bibnamefont
  {Munk}}, \bibinfo {author} {\bibfnamefont {W.~C.}\ \bibnamefont {O'Reilly}},
  \ and\ \bibinfo {author} {\bibfnamefont {J.~L.}\ \bibnamefont {Reid}},\
  }\href {\doibase 10.1175/1520-0485(1988)018<1876:ABSTER>2.0.CO;2} {\bibfield
  {journal} {\bibinfo  {journal} {J. Phys. Oceanography}\ }\textbf {\bibinfo
  {volume} {18}},\ \bibinfo {pages} {1876} (\bibinfo {year}
  {1988})}\BibitemShut {NoStop}%
\bibitem [{\citenamefont {Makarov}\ \emph {et~al.}(2009)\citenamefont
  {Makarov}, \citenamefont {Prants}, \citenamefont {Virovlyansky},\ and\
  \citenamefont {Zaslavsky}}]{RayWave}%
  \BibitemOpen
  \bibfield  {author} {\bibinfo {author} {\bibfnamefont {D.}~\bibnamefont
  {Makarov}}, \bibinfo {author} {\bibfnamefont {S.}~\bibnamefont {Prants}},
  \bibinfo {author} {\bibfnamefont {A.}~\bibnamefont {Virovlyansky}}, \ and\
  \bibinfo {author} {\bibfnamefont {G.}~\bibnamefont {Zaslavsky}},\ }\href@noop
  {} {\emph {\bibinfo {title} {{Ray and wave chaos in ocean acoustics: chaos in
  waveguides}}}},\ {Series on complexity, nonlinearity and chaos}\ (\bibinfo
  {publisher} {World Scientific},\ \bibinfo {address} {Singapore},\ \bibinfo
  {year} {2009})\ p.\ \bibinfo {pages} {388}\BibitemShut {NoStop}%
\bibitem [{\citenamefont {Smith}\ \emph {et~al.}(1992)\citenamefont {Smith},
  \citenamefont {Brown},\ and\ \citenamefont {Tappert}}]{Smith1}%
  \BibitemOpen
  \bibfield  {author} {\bibinfo {author} {\bibfnamefont {K.~B.}\ \bibnamefont
  {Smith}}, \bibinfo {author} {\bibfnamefont {M.~G.}\ \bibnamefont {Brown}}, \
  and\ \bibinfo {author} {\bibfnamefont {F.~D.}\ \bibnamefont {Tappert}},\
  }\href {\doibase 10.1121/1.403677} {\bibfield  {journal} {\bibinfo  {journal}
  {J. Acoust. Soc. Am.}\ }\textbf {\bibinfo {volume} {91}},\ \bibinfo {pages}
  {1939} (\bibinfo {year} {1992})}\BibitemShut {NoStop}%
\bibitem [{\citenamefont {Smirnov}\ \emph {et~al.}(2001)\citenamefont
  {Smirnov}, \citenamefont {Virovlyansky},\ and\ \citenamefont
  {Zaslavsky}}]{Viro01}%
  \BibitemOpen
  \bibfield  {author} {\bibinfo {author} {\bibfnamefont {I.~P.}\ \bibnamefont
  {Smirnov}}, \bibinfo {author} {\bibfnamefont {A.~L.}\ \bibnamefont
  {Virovlyansky}}, \ and\ \bibinfo {author} {\bibfnamefont {G.~M.}\
  \bibnamefont {Zaslavsky}},\ }\href {\doibase 10.1103/PhysRevE.64.036221}
  {\bibfield  {journal} {\bibinfo  {journal} {Phys. Rev. E}\ }\textbf {\bibinfo
  {volume} {64}},\ \bibinfo {pages} {036221} (\bibinfo {year}
  {2001})}\BibitemShut {NoStop}%
\bibitem [{\citenamefont {Brown}\ \emph {et~al.}(2003)\citenamefont {Brown},
  \citenamefont {Colosi}, \citenamefont {Tomsovic}, \citenamefont
  {Virovlyansky}, \citenamefont {Wolfson},\ and\ \citenamefont
  {Zaslavsky}}]{Review03}%
  \BibitemOpen
  \bibfield  {author} {\bibinfo {author} {\bibfnamefont {M.~G.}\ \bibnamefont
  {Brown}}, \bibinfo {author} {\bibfnamefont {J.~A.}\ \bibnamefont {Colosi}},
  \bibinfo {author} {\bibfnamefont {S.}~\bibnamefont {Tomsovic}}, \bibinfo
  {author} {\bibfnamefont {A.~L.}\ \bibnamefont {Virovlyansky}}, \bibinfo
  {author} {\bibfnamefont {M.~A.}\ \bibnamefont {Wolfson}}, \ and\ \bibinfo
  {author} {\bibfnamefont {G.~M.}\ \bibnamefont {Zaslavsky}},\ }\href {\doibase
  10.1121/1.1563670} {\bibfield  {journal} {\bibinfo  {journal} {J. Acoust.
  Soc. Am.}\ }\textbf {\bibinfo {volume} {113}},\ \bibinfo {pages} {2533}
  (\bibinfo {year} {2003})}\BibitemShut {NoStop}%
\bibitem [{\citenamefont {Makarov}\ \emph {et~al.}(2004)\citenamefont
  {Makarov}, \citenamefont {Uleysky},\ and\ \citenamefont {Prants}}]{Chaos}%
  \BibitemOpen
  \bibfield  {author} {\bibinfo {author} {\bibfnamefont {D.~V.}\ \bibnamefont
  {Makarov}}, \bibinfo {author} {\bibfnamefont {M.~Y.}\ \bibnamefont
  {Uleysky}}, \ and\ \bibinfo {author} {\bibfnamefont {S.~V.}\ \bibnamefont
  {Prants}},\ }\href {\doibase 10.1063/1.1626392} {\bibfield  {journal}
  {\bibinfo  {journal} {Chaos}\ }\textbf {\bibinfo {volume} {14}},\ \bibinfo
  {pages} {79} (\bibinfo {year} {2004})}\BibitemShut {NoStop}%
\bibitem [{\citenamefont {Beron-Vera}\ and\ \citenamefont
  {Brown}(2004)}]{BV04}%
  \BibitemOpen
  \bibfield  {author} {\bibinfo {author} {\bibfnamefont {F.~J.}\ \bibnamefont
  {Beron-Vera}}\ and\ \bibinfo {author} {\bibfnamefont {M.~G.}\ \bibnamefont
  {Brown}},\ }\href {\doibase 10.1121/1.1648320} {\bibfield  {journal}
  {\bibinfo  {journal} {J. Acoust. Soc. Am.}\ }\textbf {\bibinfo {volume}
  {115}},\ \bibinfo {pages} {1068} (\bibinfo {year} {2004})}\BibitemShut
  {NoStop}%
\bibitem [{\citenamefont {Virovlyansky}\ and\ \citenamefont
  {Zaslavsky}(1999)}]{Viro99}%
  \BibitemOpen
  \bibfield  {author} {\bibinfo {author} {\bibfnamefont {A.~L.}\ \bibnamefont
  {Virovlyansky}}\ and\ \bibinfo {author} {\bibfnamefont {G.~M.}\ \bibnamefont
  {Zaslavsky}},\ }\href {\doibase 10.1103/PhysRevE.59.1656} {\bibfield
  {journal} {\bibinfo  {journal} {Phys. Rev. E}\ }\textbf {\bibinfo {volume}
  {59}},\ \bibinfo {pages} {1656} (\bibinfo {year} {1999})}\BibitemShut
  {NoStop}%
\bibitem [{\citenamefont {Smirnov}\ \emph {et~al.}(2004)\citenamefont
  {Smirnov}, \citenamefont {Virovlyansky},\ and\ \citenamefont
  {Zaslavsky}}]{Viro04}%
  \BibitemOpen
  \bibfield  {author} {\bibinfo {author} {\bibfnamefont {I.~P.}\ \bibnamefont
  {Smirnov}}, \bibinfo {author} {\bibfnamefont {A.~L.}\ \bibnamefont
  {Virovlyansky}}, \ and\ \bibinfo {author} {\bibfnamefont {G.~M.}\
  \bibnamefont {Zaslavsky}},\ }\href {\doibase 10.1063/1.1737271} {\bibfield
  {journal} {\bibinfo  {journal} {Chaos}\ }\textbf {\bibinfo {volume} {14}},\
  \bibinfo {pages} {317} (\bibinfo {year} {2004})}\BibitemShut {NoStop}%
\bibitem [{\citenamefont {Smirnov}\ \emph {et~al.}(2005)\citenamefont
  {Smirnov}, \citenamefont {Virovlyansky}, \citenamefont {Edelman},\ and\
  \citenamefont {Zaslavsky}}]{Viro05}%
  \BibitemOpen
  \bibfield  {author} {\bibinfo {author} {\bibfnamefont {I.~P.}\ \bibnamefont
  {Smirnov}}, \bibinfo {author} {\bibfnamefont {A.~L.}\ \bibnamefont
  {Virovlyansky}}, \bibinfo {author} {\bibfnamefont {M.}~\bibnamefont
  {Edelman}}, \ and\ \bibinfo {author} {\bibfnamefont {G.~M.}\ \bibnamefont
  {Zaslavsky}},\ }\href {\doibase 10.1103/PhysRevE.72.026206} {\bibfield
  {journal} {\bibinfo  {journal} {Phys. Rev. E}\ }\textbf {\bibinfo {volume}
  {72}},\ \bibinfo {pages} {026206} (\bibinfo {year} {2005})}\BibitemShut
  {NoStop}%
\bibitem [{\citenamefont {Kon'kov}\ \emph {et~al.}(2007)\citenamefont
  {Kon'kov}, \citenamefont {Makarov}, \citenamefont {Sosedko},\ and\
  \citenamefont {Uleysky}}]{PRE76}%
  \BibitemOpen
  \bibfield  {author} {\bibinfo {author} {\bibfnamefont {L.~E.}\ \bibnamefont
  {Kon'kov}}, \bibinfo {author} {\bibfnamefont {D.~V.}\ \bibnamefont
  {Makarov}}, \bibinfo {author} {\bibfnamefont {E.~V.}\ \bibnamefont
  {Sosedko}}, \ and\ \bibinfo {author} {\bibfnamefont {M.~Y.}\ \bibnamefont
  {Uleysky}},\ }\href {\doibase 10.1103/PhysRevE.76.056212} {\bibfield
  {journal} {\bibinfo  {journal} {Phys. Rev. E}\ }\textbf {\bibinfo {volume}
  {76}},\ \bibinfo {pages} {056212} (\bibinfo {year} {2007})}\BibitemShut
  {NoStop}%
\bibitem [{\citenamefont {Makarov}\ \emph {et~al.}(2008)\citenamefont
  {Makarov}, \citenamefont {Kon'kov},\ and\ \citenamefont
  {Uleysky}}]{Acoust08}%
  \BibitemOpen
  \bibfield  {author} {\bibinfo {author} {\bibfnamefont {D.~V.}\ \bibnamefont
  {Makarov}}, \bibinfo {author} {\bibfnamefont {L.~E.}\ \bibnamefont
  {Kon'kov}}, \ and\ \bibinfo {author} {\bibfnamefont {M.~Y.}\ \bibnamefont
  {Uleysky}},\ }\href {\doibase 10.1134/S1063771008030147} {\bibfield
  {journal} {\bibinfo  {journal} {Acoust. Phys.}\ }\textbf {\bibinfo {volume}
  {54}},\ \bibinfo {pages} {382} (\bibinfo {year} {2008})}\BibitemShut
  {NoStop}%
\bibitem [{\citenamefont {Virovlyansky}\ \emph {et~al.}(2012)\citenamefont
  {Virovlyansky}, \citenamefont {Makarov},\ and\ \citenamefont {Prants}}]{UFN}%
  \BibitemOpen
  \bibfield  {author} {\bibinfo {author} {\bibfnamefont {A.~L.}\ \bibnamefont
  {Virovlyansky}}, \bibinfo {author} {\bibfnamefont {D.~V.}\ \bibnamefont
  {Makarov}}, \ and\ \bibinfo {author} {\bibfnamefont {S.~V.}\ \bibnamefont
  {Prants}},\ }\href@noop {} {\bibfield  {journal} {\bibinfo  {journal} {Phys.
  Usp.}\ }\textbf {\bibinfo {volume} {55}} (\bibinfo {year}
  {2012})}\BibitemShut {NoStop}%
\bibitem [{\citenamefont {Hegewisch}\ \emph {et~al.}(2005)\citenamefont
  {Hegewisch}, \citenamefont {Cerruti},\ and\ \citenamefont {Tomsovic}}]{Hege}%
  \BibitemOpen
  \bibfield  {author} {\bibinfo {author} {\bibfnamefont {K.~C.}\ \bibnamefont
  {Hegewisch}}, \bibinfo {author} {\bibfnamefont {N.~R.}\ \bibnamefont
  {Cerruti}}, \ and\ \bibinfo {author} {\bibfnamefont {S.}~\bibnamefont
  {Tomsovic}},\ }\href {\doibase 10.1121/1.1854842} {\bibfield  {journal}
  {\bibinfo  {journal} {J. Acoust. Soc. Am.}\ }\textbf {\bibinfo {volume}
  {117}},\ \bibinfo {pages} {1582} (\bibinfo {year} {2005})}\BibitemShut
  {NoStop}%
\bibitem [{\citenamefont {Munk}\ and\ \citenamefont {Wunsch}(1979)}]{MW79}%
  \BibitemOpen
  \bibfield  {author} {\bibinfo {author} {\bibfnamefont {W.}~\bibnamefont
  {Munk}}\ and\ \bibinfo {author} {\bibfnamefont {C.}~\bibnamefont {Wunsch}},\
  }\href {\doibase 10.1016/0198-0149(79)90073-6} {\bibfield  {journal}
  {\bibinfo  {journal} {Deep Sea Res. Part A.}\ }\textbf {\bibinfo {volume}
  {26}},\ \bibinfo {pages} {123} (\bibinfo {year} {1979})}\BibitemShut
  {NoStop}%
\bibitem [{\citenamefont {Tappert}\ and\ \citenamefont {Tang}(1996)}]{Tap}%
  \BibitemOpen
  \bibfield  {author} {\bibinfo {author} {\bibfnamefont {F.~D.}\ \bibnamefont
  {Tappert}}\ and\ \bibinfo {author} {\bibfnamefont {X.}~\bibnamefont {Tang}},\
  }\href {\doibase 10.1121/1.414502} {\bibfield  {journal} {\bibinfo  {journal}
  {J. Acoust. Soc. Am.}\ }\textbf {\bibinfo {volume} {99}},\ \bibinfo {pages}
  {185} (\bibinfo {year} {1996})}\BibitemShut {NoStop}%
\bibitem [{\citenamefont {Zaslavsky}(2007)}]{Zas}%
  \BibitemOpen
  \bibfield  {author} {\bibinfo {author} {\bibfnamefont {G.~M.}\ \bibnamefont
  {Zaslavsky}},\ }\href@noop {} {\emph {\bibinfo {title} {{The physics of chaos
  in Hamiltonian systems}}}}\ (\bibinfo  {publisher} {Imperial College Press},\
  \bibinfo {address} {London},\ \bibinfo {year} {2007})\ p.\ \bibinfo {pages}
  {328}\BibitemShut {NoStop}%
\bibitem [{\citenamefont {Dozier}\ and\ \citenamefont
  {Tappert}(1978{\natexlab{a}})}]{DozierI}%
  \BibitemOpen
  \bibfield  {author} {\bibinfo {author} {\bibfnamefont {L.~B.}\ \bibnamefont
  {Dozier}}\ and\ \bibinfo {author} {\bibfnamefont {F.~D.}\ \bibnamefont
  {Tappert}},\ }\href {\doibase 10.1121/1.381746} {\bibfield  {journal}
  {\bibinfo  {journal} {J. Acoust. Soc. Am.}\ }\textbf {\bibinfo {volume}
  {63}},\ \bibinfo {pages} {353} (\bibinfo {year}
  {1978}{\natexlab{a}})}\BibitemShut {NoStop}%
\bibitem [{\citenamefont {Colosi}\ and\ \citenamefont
  {Morozov}(2009)}]{ColosiMorozov}%
  \BibitemOpen
  \bibfield  {author} {\bibinfo {author} {\bibfnamefont {J.~A.}\ \bibnamefont
  {Colosi}}\ and\ \bibinfo {author} {\bibfnamefont {A.~K.}\ \bibnamefont
  {Morozov}},\ }\href {\doibase 10.1121/1.3158818} {\bibfield  {journal}
  {\bibinfo  {journal} {J. Acoust. Soc. Am.}\ }\textbf {\bibinfo {volume}
  {126}},\ \bibinfo {pages} {1026} (\bibinfo {year} {2009})}\BibitemShut
  {NoStop}%
\bibitem [{\citenamefont {Wolfson}\ and\ \citenamefont
  {Tomsovic}(2001)}]{WT01}%
  \BibitemOpen
  \bibfield  {author} {\bibinfo {author} {\bibfnamefont {M.~A.}\ \bibnamefont
  {Wolfson}}\ and\ \bibinfo {author} {\bibfnamefont {S.}~\bibnamefont
  {Tomsovic}},\ }\href {\doibase 10.1121/1.1362685} {\bibfield  {journal}
  {\bibinfo  {journal} {J. Acoust. Soc. Am.}\ }\textbf {\bibinfo {volume}
  {109}},\ \bibinfo {pages} {2693} (\bibinfo {year} {2001})}\BibitemShut
  {NoStop}%
\bibitem [{\citenamefont {Makarov}\ \emph {et~al.}(2006)\citenamefont
  {Makarov}, \citenamefont {Uleysky}, \citenamefont {Budyansky},\ and\
  \citenamefont {Prants}}]{PRE73}%
  \BibitemOpen
  \bibfield  {author} {\bibinfo {author} {\bibfnamefont {D.~V.}\ \bibnamefont
  {Makarov}}, \bibinfo {author} {\bibfnamefont {M.~Y.}\ \bibnamefont
  {Uleysky}}, \bibinfo {author} {\bibfnamefont {M.~V.}\ \bibnamefont
  {Budyansky}}, \ and\ \bibinfo {author} {\bibfnamefont {S.~V.}\ \bibnamefont
  {Prants}},\ }\href {\doibase 10.1103/PhysRevE.73.066210} {\bibfield
  {journal} {\bibinfo  {journal} {Phys. Rev. E}\ }\textbf {\bibinfo {volume}
  {73}},\ \bibinfo {pages} {066210} (\bibinfo {year} {2006})}\BibitemShut
  {NoStop}%
\bibitem [{\citenamefont {Makarov}\ \emph {et~al.}(2010)\citenamefont
  {Makarov}, \citenamefont {Kon'kov},\ and\ \citenamefont {Uleysky}}]{Arxiv}%
  \BibitemOpen
  \bibfield  {author} {\bibinfo {author} {\bibfnamefont {D.~V.}\ \bibnamefont
  {Makarov}}, \bibinfo {author} {\bibfnamefont {L.~E.}\ \bibnamefont
  {Kon'kov}}, \ and\ \bibinfo {author} {\bibfnamefont {M.~Y.}\ \bibnamefont
  {Uleysky}},\ }\href@noop {} {\bibfield  {journal} {\bibinfo  {journal} {ArXiv
  e-prints}\ } (\bibinfo {year} {2010})},\ \Eprint
  {http://arxiv.org/abs/1008.3037} {arXiv:1008.3037 [nlin.CD]} \BibitemShut
  {NoStop}%
\bibitem [{\citenamefont {Hegewisch}\ and\ \citenamefont
  {Tomsovic}(2012)}]{Tomc11}%
  \BibitemOpen
  \bibfield  {author} {\bibinfo {author} {\bibfnamefont {K.~C.}\ \bibnamefont
  {Hegewisch}}\ and\ \bibinfo {author} {\bibfnamefont {S.}~\bibnamefont
  {Tomsovic}},\ }\href {\doibase 10.1209/0295-5075/97/34002} {\bibfield
  {journal} {\bibinfo  {journal} {Europhys. Lett.}\ }\textbf {\bibinfo {volume}
  {97}},\ \bibinfo {pages} {34002} (\bibinfo {year} {2012})}\BibitemShut
  {NoStop}%
\bibitem [{\citenamefont {St{\"{o}}ckmann}(2007)}]{Stockman}%
  \BibitemOpen
  \bibfield  {author} {\bibinfo {author} {\bibfnamefont {H.~J.}\ \bibnamefont
  {St{\"{o}}ckmann}},\ }\href {\doibase 10.2277/0521027152} {\emph {\bibinfo
  {title} {{Quantum Chaos: An Introduction}}}}\ (\bibinfo  {publisher}
  {Cambridge University Press},\ \bibinfo {address} {Cambridge},\ \bibinfo
  {year} {2007})\ p.\ \bibinfo {pages} {384}\BibitemShut {NoStop}%
\bibitem [{\citenamefont {Kolovsky}(1997)}]{Kol97}%
  \BibitemOpen
  \bibfield  {author} {\bibinfo {author} {\bibfnamefont {A.~R.}\ \bibnamefont
  {Kolovsky}},\ }\href {\doibase 10.1103/PhysRevE.56.2261} {\bibfield
  {journal} {\bibinfo  {journal} {Phys. Rev. E}\ }\textbf {\bibinfo {volume}
  {56}},\ \bibinfo {pages} {2261} (\bibinfo {year} {1997})}\BibitemShut
  {NoStop}%
\bibitem [{\citenamefont {Bezotvetnykh}\ \emph {et~al.}(2009)\citenamefont
  {Bezotvetnykh}, \citenamefont {Burenin}, \citenamefont {Morgunov},\ and\
  \citenamefont {Polovinka}}]{Bez09}%
  \BibitemOpen
  \bibfield  {author} {\bibinfo {author} {\bibfnamefont {V.}~\bibnamefont
  {Bezotvetnykh}}, \bibinfo {author} {\bibfnamefont {A.}~\bibnamefont
  {Burenin}}, \bibinfo {author} {\bibfnamefont {Y.}~\bibnamefont {Morgunov}}, \
  and\ \bibinfo {author} {\bibfnamefont {Y.}~\bibnamefont {Polovinka}},\ }\href
  {\doibase 10.1134/S1063771009030130} {\bibfield  {journal} {\bibinfo
  {journal} {Acoust. Phys.}\ }\textbf {\bibinfo {volume} {55}},\ \bibinfo
  {pages} {376} (\bibinfo {year} {2009})}\BibitemShut {NoStop}%
\bibitem [{\citenamefont {Spindel}\ \emph {et~al.}(2003)\citenamefont
  {Spindel}, \citenamefont {Na}, \citenamefont {Dahl}, \citenamefont {Oh},
  \citenamefont {Eggen}, \citenamefont {Kim}, \citenamefont {Akulichev},\ and\
  \citenamefont {Morgunov}}]{IEEE}%
  \BibitemOpen
  \bibfield  {author} {\bibinfo {author} {\bibfnamefont {R.~C.}\ \bibnamefont
  {Spindel}}, \bibinfo {author} {\bibfnamefont {J.}~\bibnamefont {Na}},
  \bibinfo {author} {\bibfnamefont {P.~H.}\ \bibnamefont {Dahl}}, \bibinfo
  {author} {\bibfnamefont {S.}~\bibnamefont {Oh}}, \bibinfo {author}
  {\bibfnamefont {C.}~\bibnamefont {Eggen}}, \bibinfo {author} {\bibfnamefont
  {Y.~G.}\ \bibnamefont {Kim}}, \bibinfo {author} {\bibfnamefont {V.~A.}\
  \bibnamefont {Akulichev}}, \ and\ \bibinfo {author} {\bibfnamefont {Y.~N.}\
  \bibnamefont {Morgunov}},\ }\href {\doibase 10.1109/JOE.2003.811896}
  {\bibfield  {journal} {\bibinfo  {journal} {IEEE J. Ocean. Engin.}\ }\textbf
  {\bibinfo {volume} {28}},\ \bibinfo {pages} {297} (\bibinfo {year}
  {2003})}\BibitemShut {NoStop}%
\bibitem [{Dat()}]{Database}%
  \BibitemOpen
  \href@noop {} {\enquote {\bibinfo {title} {{Oceanography and marine
  environment of the Far Eastern Region of Russia (proj. leader
  Rostov~I.~D.)}},}\ }\bibinfo {howpublished}
  {\url{http://www.pacificinfo.ru/en}}\BibitemShut {NoStop}%
\bibitem [{\citenamefont {Brekhovskikh}\ \emph {et~al.}(1990)\citenamefont
  {Brekhovskikh}, \citenamefont {Goncharov}, \citenamefont {Dremuchev},
  \citenamefont {Kurtepov}, \citenamefont {Selivanov},\ and\ \citenamefont
  {Chepurin}}]{Brekh90}%
  \BibitemOpen
  \bibfield  {author} {\bibinfo {author} {\bibfnamefont {L.~M.}\ \bibnamefont
  {Brekhovskikh}}, \bibinfo {author} {\bibfnamefont {V.~V.}\ \bibnamefont
  {Goncharov}}, \bibinfo {author} {\bibfnamefont {S.~A.}\ \bibnamefont
  {Dremuchev}}, \bibinfo {author} {\bibfnamefont {V.~M.}\ \bibnamefont
  {Kurtepov}}, \bibinfo {author} {\bibfnamefont {V.~G.}\ \bibnamefont
  {Selivanov}}, \ and\ \bibinfo {author} {\bibfnamefont {Y.~A.}\ \bibnamefont
  {Chepurin}},\ }\href@noop {} {\bibfield  {journal} {\bibinfo  {journal} {Sov.
  Phys. Acoust.}\ }\textbf {\bibinfo {volume} {36}},\ \bibinfo {pages} {461}
  (\bibinfo {year} {1990})}\BibitemShut {NoStop}%
\bibitem [{\citenamefont {Smirnov}\ \emph {et~al.}(1999)\citenamefont
  {Smirnov}, \citenamefont {Caruthers},\ and\ \citenamefont
  {Khil'ko}}]{Caruth}%
  \BibitemOpen
  \bibfield  {author} {\bibinfo {author} {\bibfnamefont {I.}~\bibnamefont
  {Smirnov}}, \bibinfo {author} {\bibfnamefont {J.}~\bibnamefont {Caruthers}},
  \ and\ \bibinfo {author} {\bibfnamefont {A.}~\bibnamefont {Khil'ko}},\ }\href
  {\doibase 10.1007/BF02677099} {\bibfield  {journal} {\bibinfo  {journal}
  {Radiophys. Quantum Electron.}\ }\textbf {\bibinfo {volume} {42}},\ \bibinfo
  {pages} {864} (\bibinfo {year} {1999})}\BibitemShut {NoStop}%
\bibitem [{\citenamefont {Morozov}\ and\ \citenamefont
  {Colosi}(2005)}]{MorCol}%
  \BibitemOpen
  \bibfield  {author} {\bibinfo {author} {\bibfnamefont {A.~K.}\ \bibnamefont
  {Morozov}}\ and\ \bibinfo {author} {\bibfnamefont {J.~A.}\ \bibnamefont
  {Colosi}},\ }\href {\doibase 10.1121/1.1854571} {\bibfield  {journal}
  {\bibinfo  {journal} {J. Acoust. Soc. Am.}\ }\textbf {\bibinfo {volume}
  {117}},\ \bibinfo {pages} {1611} (\bibinfo {year} {2005})}\BibitemShut
  {NoStop}%
\bibitem [{\citenamefont {Petukhov}(2009)}]{Petukhov}%
  \BibitemOpen
  \bibfield  {author} {\bibinfo {author} {\bibfnamefont {Y.}~\bibnamefont
  {Petukhov}},\ }\href {\doibase 10.1134/S106377100906013X} {\bibfield
  {journal} {\bibinfo  {journal} {Acoust. Phys.}\ }\textbf {\bibinfo {volume}
  {55}},\ \bibinfo {pages} {785} (\bibinfo {year} {2009})}\BibitemShut
  {NoStop}%
\bibitem [{\citenamefont {Colosi}\ and\ \citenamefont {Brown}(1998)}]{CB}%
  \BibitemOpen
  \bibfield  {author} {\bibinfo {author} {\bibfnamefont {J.~A.}\ \bibnamefont
  {Colosi}}\ and\ \bibinfo {author} {\bibfnamefont {M.~G.}\ \bibnamefont
  {Brown}},\ }\href {\doibase 10.1121/1.421381} {\bibfield  {journal} {\bibinfo
   {journal} {J. Acoust. Soc. Am.}\ }\textbf {\bibinfo {volume} {103}},\
  \bibinfo {pages} {2232} (\bibinfo {year} {1998})}\BibitemShut {NoStop}%
\bibitem [{\citenamefont {LeBlanc}\ and\ \citenamefont
  {Middleton}(1980)}]{LeBlanc}%
  \BibitemOpen
  \bibfield  {author} {\bibinfo {author} {\bibfnamefont {L.~R.}\ \bibnamefont
  {LeBlanc}}\ and\ \bibinfo {author} {\bibfnamefont {F.~H.}\ \bibnamefont
  {Middleton}},\ }\href {\doibase 10.1121/1.384448} {\bibfield  {journal}
  {\bibinfo  {journal} {J. Acoust. Soc. Am.}\ }\textbf {\bibinfo {volume}
  {67}},\ \bibinfo {pages} {2055} (\bibinfo {year} {1980})}\BibitemShut
  {NoStop}%
\bibitem [{\citenamefont {Dozier}\ and\ \citenamefont
  {Tappert}(1978{\natexlab{b}})}]{DozierII}%
  \BibitemOpen
  \bibfield  {author} {\bibinfo {author} {\bibfnamefont {L.~B.}\ \bibnamefont
  {Dozier}}\ and\ \bibinfo {author} {\bibfnamefont {F.~D.}\ \bibnamefont
  {Tappert}},\ }\href {\doibase 10.1121/1.382005} {\bibfield  {journal}
  {\bibinfo  {journal} {J. Acoust. Soc. Am.}\ }\textbf {\bibinfo {volume}
  {64}},\ \bibinfo {pages} {533} (\bibinfo {year}
  {1978}{\natexlab{b}})}\BibitemShut {NoStop}%
\bibitem [{\citenamefont {Mallick}\ and\ \citenamefont
  {Marcq}(2002)}]{Mallick}%
  \BibitemOpen
  \bibfield  {author} {\bibinfo {author} {\bibfnamefont {K.}~\bibnamefont
  {Mallick}}\ and\ \bibinfo {author} {\bibfnamefont {P.}~\bibnamefont
  {Marcq}},\ }\href {\doibase 10.1103/PhysRevE.66.041113} {\bibfield  {journal}
  {\bibinfo  {journal} {Phys. Rev. E}\ }\textbf {\bibinfo {volume} {66}},\
  \bibinfo {pages} {041113} (\bibinfo {year} {2002})}\BibitemShut {NoStop}%
\bibitem [{\citenamefont {Landau}\ and\ \citenamefont {Lifshitz}(1977)}]{LL3}%
  \BibitemOpen
  \bibfield  {author} {\bibinfo {author} {\bibfnamefont {L.~D.}\ \bibnamefont
  {Landau}}\ and\ \bibinfo {author} {\bibfnamefont {E.~M.}\ \bibnamefont
  {Lifshitz}},\ }\href@noop {} {\emph {\bibinfo {title} {{Course of theoretical
  physics.}}}},\ Vol.\ \bibinfo {volume} {3: Quantum mechanics. Nonrelativistic
  theory}\ (\bibinfo  {publisher} {Pergamon Press},\ \bibinfo {address}
  {Oxford},\ \bibinfo {year} {1977})\BibitemShut {NoStop}%
\bibitem [{\citenamefont {Berry}\ and\ \citenamefont {Robnik}(1984)}]{BR}%
  \BibitemOpen
  \bibfield  {author} {\bibinfo {author} {\bibfnamefont {M.~V.}\ \bibnamefont
  {Berry}}\ and\ \bibinfo {author} {\bibfnamefont {M.}~\bibnamefont {Robnik}},\
  }\href {\doibase 10.1088/0305-4470/17/12/013} {\bibfield  {journal} {\bibinfo
   {journal} {J. Phys. A: Math. Gen.}\ }\textbf {\bibinfo {volume} {17}},\
  \bibinfo {pages} {2413} (\bibinfo {year} {1984})}\BibitemShut {NoStop}%
\bibitem [{\citenamefont {B{\"{a}}cker}\ \emph {et~al.}(2005)\citenamefont
  {B{\"{a}}cker}, \citenamefont {Ketzmerick},\ and\ \citenamefont
  {Monastra}}]{Backer}%
  \BibitemOpen
  \bibfield  {author} {\bibinfo {author} {\bibfnamefont {A.}~\bibnamefont
  {B{\"{a}}cker}}, \bibinfo {author} {\bibfnamefont {R.}~\bibnamefont
  {Ketzmerick}}, \ and\ \bibinfo {author} {\bibfnamefont {A.~G.}\ \bibnamefont
  {Monastra}},\ }\href {\doibase 10.1103/PhysRevLett.94.054102} {\bibfield
  {journal} {\bibinfo  {journal} {Phys. Rev. Lett.}\ }\textbf {\bibinfo
  {volume} {94}},\ \bibinfo {pages} {054102} (\bibinfo {year}
  {2005})}\BibitemShut {NoStop}%
\bibitem [{\citenamefont {Prosen}\ and\ \citenamefont {Robnik}(1994)}]{PR}%
  \BibitemOpen
  \bibfield  {author} {\bibinfo {author} {\bibfnamefont {T.}~\bibnamefont
  {Prosen}}\ and\ \bibinfo {author} {\bibfnamefont {M.}~\bibnamefont
  {Robnik}},\ }\href {\doibase 10.1088/0305-4470/27/24/017} {\bibfield
  {journal} {\bibinfo  {journal} {J. Phys. A: Math. Gen.}\ }\textbf {\bibinfo
  {volume} {27}},\ \bibinfo {pages} {8059} (\bibinfo {year}
  {1994})}\BibitemShut {NoStop}%
\bibitem [{\citenamefont {Varga}\ and\ \citenamefont {Pipek}(2003)}]{Varga}%
  \BibitemOpen
  \bibfield  {author} {\bibinfo {author} {\bibfnamefont {I.}~\bibnamefont
  {Varga}}\ and\ \bibinfo {author} {\bibfnamefont {J.}~\bibnamefont {Pipek}},\
  }\href {\doibase 10.1103/PhysRevE.68.026202} {\bibfield  {journal} {\bibinfo
  {journal} {Phys. Rev. E}\ }\textbf {\bibinfo {volume} {68}},\ \bibinfo
  {pages} {026202} (\bibinfo {year} {2003})}\BibitemShut {NoStop}%
\bibitem [{\citenamefont {Makarov}\ and\ \citenamefont {Uleysky}(2006)}]{JPA}%
  \BibitemOpen
  \bibfield  {author} {\bibinfo {author} {\bibfnamefont {D.}~\bibnamefont
  {Makarov}}\ and\ \bibinfo {author} {\bibfnamefont {M.}~\bibnamefont
  {Uleysky}},\ }\href {\doibase 10.1088/0305-4470/39/3/003} {\bibfield
  {journal} {\bibinfo  {journal} {J. Phys. A: Math. Gen.}\ }\textbf {\bibinfo
  {volume} {39}},\ \bibinfo {pages} {489} (\bibinfo {year} {2006})}\BibitemShut
  {NoStop}%
\bibitem [{\citenamefont {Gan}\ \emph {et~al.}(2010)\citenamefont {Gan},
  \citenamefont {Wang},\ and\ \citenamefont {Perc}}]{Gan1}%
  \BibitemOpen
  \bibfield  {author} {\bibinfo {author} {\bibfnamefont {C.}~\bibnamefont
  {Gan}}, \bibinfo {author} {\bibfnamefont {Q.}~\bibnamefont {Wang}}, \ and\
  \bibinfo {author} {\bibfnamefont {M.}~\bibnamefont {Perc}},\ }\href {\doibase
  10.1088/1751-8113/43/12/125102} {\bibfield  {journal} {\bibinfo  {journal}
  {J. Physics A: Math. Theor.}\ }\textbf {\bibinfo {volume} {43}},\ \bibinfo
  {pages} {125102} (\bibinfo {year} {2010})}\BibitemShut {NoStop}%
\bibitem [{\citenamefont {Gan}\ and\ \citenamefont {Lei}(2011)}]{Gan2}%
  \BibitemOpen
  \bibfield  {author} {\bibinfo {author} {\bibfnamefont {C.}~\bibnamefont
  {Gan}}\ and\ \bibinfo {author} {\bibfnamefont {H.}~\bibnamefont {Lei}},\
  }\href {\doibase 10.1016/j.jsv.2010.09.025} {\bibfield  {journal} {\bibinfo
  {journal} {J. Sound Vibr.}\ }\textbf {\bibinfo {volume} {330}},\ \bibinfo
  {pages} {2174} (\bibinfo {year} {2011})}\BibitemShut {NoStop}%
\bibitem [{\citenamefont {Rypina}\ \emph {et~al.}(2007)\citenamefont {Rypina},
  \citenamefont {Brown}, \citenamefont {Beron-Vera}, \citenamefont {{Ko\c
  cak}}, \citenamefont {Olascoaga},\ and\ \citenamefont
  {Udovydchenkov}}]{Rypina}%
  \BibitemOpen
  \bibfield  {author} {\bibinfo {author} {\bibfnamefont {I.~I.}\ \bibnamefont
  {Rypina}}, \bibinfo {author} {\bibfnamefont {M.~G.}\ \bibnamefont {Brown}},
  \bibinfo {author} {\bibfnamefont {F.~J.}\ \bibnamefont {Beron-Vera}},
  \bibinfo {author} {\bibfnamefont {H.}~\bibnamefont {{Ko\c cak}}}, \bibinfo
  {author} {\bibfnamefont {M.~J.}\ \bibnamefont {Olascoaga}}, \ and\ \bibinfo
  {author} {\bibfnamefont {I.~A.}\ \bibnamefont {Udovydchenkov}},\ }\href
  {\doibase 10.1103/PhysRevLett.98.104102} {\bibfield  {journal} {\bibinfo
  {journal} {Phys. Rev. Lett.}\ }\textbf {\bibinfo {volume} {98}},\ \bibinfo
  {pages} {104102} (\bibinfo {year} {2007})}\BibitemShut {NoStop}%
\bibitem [{\citenamefont {Spiesberger}\ and\ \citenamefont
  {Tappert}(1996)}]{Kaneohe}%
  \BibitemOpen
  \bibfield  {author} {\bibinfo {author} {\bibfnamefont {J.~L.}\ \bibnamefont
  {Spiesberger}}\ and\ \bibinfo {author} {\bibfnamefont {F.~D.}\ \bibnamefont
  {Tappert}},\ }\href {\doibase 10.1121/1.414501} {\bibfield  {journal}
  {\bibinfo  {journal} {J. Acoust. Soc. Am.}\ }\textbf {\bibinfo {volume}
  {99}},\ \bibinfo {pages} {173} (\bibinfo {year} {1996})}\BibitemShut
  {NoStop}%
\bibitem [{\citenamefont {Worcester}\ \emph {et~al.}(1994)\citenamefont
  {Worcester}, \citenamefont {Cornuelle}, \citenamefont {Hildebrand},
  \citenamefont {{William S. Hodgkiss}}, \citenamefont {Duda}, \citenamefont
  {Boyd}, \citenamefont {Howe}, \citenamefont {Mercer},\ and\ \citenamefont
  {Spindel}}]{SLICE89}%
  \BibitemOpen
  \bibfield  {author} {\bibinfo {author} {\bibfnamefont {P.~F.}\ \bibnamefont
  {Worcester}}, \bibinfo {author} {\bibfnamefont {B.~D.}\ \bibnamefont
  {Cornuelle}}, \bibinfo {author} {\bibfnamefont {J.~A.}\ \bibnamefont
  {Hildebrand}}, \bibinfo {author} {\bibfnamefont {J.}~\bibnamefont {{William
  S. Hodgkiss}}}, \bibinfo {author} {\bibfnamefont {T.~F.}\ \bibnamefont
  {Duda}}, \bibinfo {author} {\bibfnamefont {J.}~\bibnamefont {Boyd}}, \bibinfo
  {author} {\bibfnamefont {B.~M.}\ \bibnamefont {Howe}}, \bibinfo {author}
  {\bibfnamefont {J.~A.}\ \bibnamefont {Mercer}}, \ and\ \bibinfo {author}
  {\bibfnamefont {R.~C.}\ \bibnamefont {Spindel}},\ }\href {\doibase
  10.1121/1.409977} {\bibfield  {journal} {\bibinfo  {journal} {J. Acoust. Soc.
  Am.}\ }\textbf {\bibinfo {volume} {95}},\ \bibinfo {pages} {3118} (\bibinfo
  {year} {1994})}\BibitemShut {NoStop}%
\bibitem [{\citenamefont {Worcester}\ \emph {et~al.}(1999)\citenamefont
  {Worcester}, \citenamefont {Cornuelle}, \citenamefont {Dzieciuch},
  \citenamefont {Munk}, \citenamefont {Howe}, \citenamefont {Mercer},
  \citenamefont {Spindel}, \citenamefont {Colosi}, \citenamefont {Metzger},
  \citenamefont {Birdsall},\ and\ \citenamefont {Baggeroer}}]{Worc-AET}%
  \BibitemOpen
  \bibfield  {author} {\bibinfo {author} {\bibfnamefont {P.~F.}\ \bibnamefont
  {Worcester}}, \bibinfo {author} {\bibfnamefont {B.~D.}\ \bibnamefont
  {Cornuelle}}, \bibinfo {author} {\bibfnamefont {M.~A.}\ \bibnamefont
  {Dzieciuch}}, \bibinfo {author} {\bibfnamefont {W.~H.}\ \bibnamefont {Munk}},
  \bibinfo {author} {\bibfnamefont {B.~M.}\ \bibnamefont {Howe}}, \bibinfo
  {author} {\bibfnamefont {J.~A.}\ \bibnamefont {Mercer}}, \bibinfo {author}
  {\bibfnamefont {R.~C.}\ \bibnamefont {Spindel}}, \bibinfo {author}
  {\bibfnamefont {J.~A.}\ \bibnamefont {Colosi}}, \bibinfo {author}
  {\bibfnamefont {K.}~\bibnamefont {Metzger}}, \bibinfo {author} {\bibfnamefont
  {T.~G.}\ \bibnamefont {Birdsall}}, \ and\ \bibinfo {author} {\bibfnamefont
  {A.~B.}\ \bibnamefont {Baggeroer}},\ }\href {\doibase 10.1121/1.424649}
  {\bibfield  {journal} {\bibinfo  {journal} {J. Acoust. Soc. Am.}\ }\textbf
  {\bibinfo {volume} {105}},\ \bibinfo {pages} {3185} (\bibinfo {year}
  {1999})}\BibitemShut {NoStop}%
\bibitem [{\citenamefont {Wage}\ \emph {et~al.}(2005)\citenamefont {Wage},
  \citenamefont {Dzieciuch}, \citenamefont {Worcester}, \citenamefont {Howe},\
  and\ \citenamefont {Mercer}}]{AST}%
  \BibitemOpen
  \bibfield  {author} {\bibinfo {author} {\bibfnamefont {K.~E.}\ \bibnamefont
  {Wage}}, \bibinfo {author} {\bibfnamefont {M.~A.}\ \bibnamefont {Dzieciuch}},
  \bibinfo {author} {\bibfnamefont {P.~F.}\ \bibnamefont {Worcester}}, \bibinfo
  {author} {\bibfnamefont {B.~M.}\ \bibnamefont {Howe}}, \ and\ \bibinfo
  {author} {\bibfnamefont {J.~A.}\ \bibnamefont {Mercer}},\ }\href {\doibase
  10.1121/1.1854551} {\bibfield  {journal} {\bibinfo  {journal} {J. Acoust.
  Soc. Am.}\ }\textbf {\bibinfo {volume} {117}},\ \bibinfo {pages} {1565}
  (\bibinfo {year} {2005})}\BibitemShut {NoStop}%
\bibitem [{\citenamefont {Grigorieva}\ \emph {et~al.}(2009)\citenamefont
  {Grigorieva}, \citenamefont {Fridman}, \citenamefont {Mercer}, \citenamefont
  {Andrew}, \citenamefont {Wolfson}, \citenamefont {Howe},\ and\ \citenamefont
  {Colosi}}]{LOAPEX}%
  \BibitemOpen
  \bibfield  {author} {\bibinfo {author} {\bibfnamefont {N.~S.}\ \bibnamefont
  {Grigorieva}}, \bibinfo {author} {\bibfnamefont {G.~M.}\ \bibnamefont
  {Fridman}}, \bibinfo {author} {\bibfnamefont {J.~A.}\ \bibnamefont {Mercer}},
  \bibinfo {author} {\bibfnamefont {R.~K.}\ \bibnamefont {Andrew}}, \bibinfo
  {author} {\bibfnamefont {M.~A.}\ \bibnamefont {Wolfson}}, \bibinfo {author}
  {\bibfnamefont {B.~M.}\ \bibnamefont {Howe}}, \ and\ \bibinfo {author}
  {\bibfnamefont {J.~A.}\ \bibnamefont {Colosi}},\ }\href {\doibase
  10.1121/1.3082112} {\bibfield  {journal} {\bibinfo  {journal} {J. Acoust.
  Soc. Am.}\ }\textbf {\bibinfo {volume} {125}},\ \bibinfo {pages} {1919}
  (\bibinfo {year} {2009})}\BibitemShut {NoStop}%
\bibitem [{\citenamefont {Simmen}\ \emph {et~al.}(1997)\citenamefont {Simmen},
  \citenamefont {Flatte},\ and\ \citenamefont {Wang}}]{Simmen}%
  \BibitemOpen
  \bibfield  {author} {\bibinfo {author} {\bibfnamefont {J.}~\bibnamefont
  {Simmen}}, \bibinfo {author} {\bibfnamefont {S.~M.}\ \bibnamefont {Flatte}},
  \ and\ \bibinfo {author} {\bibfnamefont {G.-Y.}\ \bibnamefont {Wang}},\
  }\href {\doibase 10.1121/1.419820} {\bibfield  {journal} {\bibinfo  {journal}
  {J. Acoust. Soc. Am.}\ }\textbf {\bibinfo {volume} {102}},\ \bibinfo {pages}
  {239} (\bibinfo {year} {1997})}\BibitemShut {NoStop}%
\bibitem [{\citenamefont {Beron-Vera}\ \emph {et~al.}(2003)\citenamefont
  {Beron-Vera}, \citenamefont {Brown}, \citenamefont {Colosi}, \citenamefont
  {Tomsovic}, \citenamefont {Virovlyansky}, \citenamefont {Wolfson},\ and\
  \citenamefont {Zaslavsky}}]{AET}%
  \BibitemOpen
  \bibfield  {author} {\bibinfo {author} {\bibfnamefont {F.~J.}\ \bibnamefont
  {Beron-Vera}}, \bibinfo {author} {\bibfnamefont {M.~G.}\ \bibnamefont
  {Brown}}, \bibinfo {author} {\bibfnamefont {J.~A.}\ \bibnamefont {Colosi}},
  \bibinfo {author} {\bibfnamefont {S.}~\bibnamefont {Tomsovic}}, \bibinfo
  {author} {\bibfnamefont {A.~L.}\ \bibnamefont {Virovlyansky}}, \bibinfo
  {author} {\bibfnamefont {M.~A.}\ \bibnamefont {Wolfson}}, \ and\ \bibinfo
  {author} {\bibfnamefont {G.~M.}\ \bibnamefont {Zaslavsky}},\ }\href {\doibase
  10.1121/1.1600724} {\bibfield  {journal} {\bibinfo  {journal} {J. Acoust.
  Soc. Am.}\ }\textbf {\bibinfo {volume} {114}},\ \bibinfo {pages} {1226}
  (\bibinfo {year} {2003})}\BibitemShut {NoStop}%
\bibitem [{\citenamefont {del Castillo-Negrete}\ \emph
  {et~al.}(1996)\citenamefont {del Castillo-Negrete}, \citenamefont {Greene},\
  and\ \citenamefont {Morrison}}]{Nontwist}%
  \BibitemOpen
  \bibfield  {author} {\bibinfo {author} {\bibfnamefont {D.}~\bibnamefont {del
  Castillo-Negrete}}, \bibinfo {author} {\bibfnamefont {J.~M.}\ \bibnamefont
  {Greene}}, \ and\ \bibinfo {author} {\bibfnamefont {P.~J.}\ \bibnamefont
  {Morrison}},\ }\href {\doibase 10.1016/0167-2789(95)00257-X} {\bibfield
  {journal} {\bibinfo  {journal} {Physica D}\ }\textbf {\bibinfo {volume}
  {91}},\ \bibinfo {pages} {1} (\bibinfo {year} {1996})}\BibitemShut {NoStop}%
\bibitem [{\citenamefont {Budyansky}\ \emph {et~al.}(2009)\citenamefont
  {Budyansky}, \citenamefont {Uleysky},\ and\ \citenamefont {Prants}}]{Bud79}%
  \BibitemOpen
  \bibfield  {author} {\bibinfo {author} {\bibfnamefont {M.~V.}\ \bibnamefont
  {Budyansky}}, \bibinfo {author} {\bibfnamefont {M.~Y.}\ \bibnamefont
  {Uleysky}}, \ and\ \bibinfo {author} {\bibfnamefont {S.~V.}\ \bibnamefont
  {Prants}},\ }\href {\doibase 10.1103/PhysRevE.79.056215} {\bibfield
  {journal} {\bibinfo  {journal} {Phys. Rev. E}\ }\textbf {\bibinfo {volume}
  {79}},\ \bibinfo {pages} {056215} (\bibinfo {year} {2009})}\BibitemShut
  {NoStop}%
\bibitem [{\citenamefont {Uleysky}\ \emph
  {et~al.}(2010{\natexlab{a}})\citenamefont {Uleysky}, \citenamefont
  {Budyansky},\ and\ \citenamefont {Prants}}]{Bud81}%
  \BibitemOpen
  \bibfield  {author} {\bibinfo {author} {\bibfnamefont {M.~Y.}\ \bibnamefont
  {Uleysky}}, \bibinfo {author} {\bibfnamefont {M.~V.}\ \bibnamefont
  {Budyansky}}, \ and\ \bibinfo {author} {\bibfnamefont {S.~V.}\ \bibnamefont
  {Prants}},\ }\href {\doibase 10.1103/PhysRevE.81.017202} {\bibfield
  {journal} {\bibinfo  {journal} {Phys. Rev. E}\ }\textbf {\bibinfo {volume}
  {81}},\ \bibinfo {pages} {017202} (\bibinfo {year}
  {2010}{\natexlab{a}})}\BibitemShut {NoStop}%
\bibitem [{\citenamefont {Uleysky}\ \emph
  {et~al.}(2010{\natexlab{b}})\citenamefont {Uleysky}, \citenamefont
  {Budyansky},\ and\ \citenamefont {Prants}}]{JETP10}%
  \BibitemOpen
  \bibfield  {author} {\bibinfo {author} {\bibfnamefont {M.}~\bibnamefont
  {Uleysky}}, \bibinfo {author} {\bibfnamefont {M.}~\bibnamefont {Budyansky}},
  \ and\ \bibinfo {author} {\bibfnamefont {S.}~\bibnamefont {Prants}},\ }\href
  {\doibase 10.1134/S1063776110120174} {\bibfield  {journal} {\bibinfo
  {journal} {J. of Exp. and Theor. Phys.}\ }\textbf {\bibinfo {volume} {111}},\
  \bibinfo {pages} {1039} (\bibinfo {year} {2010}{\natexlab{b}})}\BibitemShut
  {NoStop}%
\bibitem [{\citenamefont {{Rela{\~{n}}o}}\ \emph {et~al.}(2002)\citenamefont
  {{Rela{\~{n}}o}}, \citenamefont {G{\'{o}}mez}, \citenamefont {Molina},
  \citenamefont {Retamosa},\ and\ \citenamefont {Faleiro}}]{Relano02}%
  \BibitemOpen
  \bibfield  {author} {\bibinfo {author} {\bibfnamefont {A.}~\bibnamefont
  {{Rela{\~{n}}o}}}, \bibinfo {author} {\bibfnamefont {J.~M.~G.}\ \bibnamefont
  {G{\'{o}}mez}}, \bibinfo {author} {\bibfnamefont {R.~A.}\ \bibnamefont
  {Molina}}, \bibinfo {author} {\bibfnamefont {J.}~\bibnamefont {Retamosa}}, \
  and\ \bibinfo {author} {\bibfnamefont {E.}~\bibnamefont {Faleiro}},\ }\href
  {\doibase 10.1103/PhysRevLett.89.244102} {\bibfield  {journal} {\bibinfo
  {journal} {Phys. Rev. Lett.}\ }\textbf {\bibinfo {volume} {89}},\ \bibinfo
  {pages} {244102} (\bibinfo {year} {2002})}\BibitemShut {NoStop}%
\bibitem [{\citenamefont {{Rela{\~{n}}o}}(2008)}]{Relano08}%
  \BibitemOpen
  \bibfield  {author} {\bibinfo {author} {\bibfnamefont {A.}~\bibnamefont
  {{Rela{\~{n}}o}}},\ }\href {\doibase 10.1103/PhysRevLett.100.224101}
  {\bibfield  {journal} {\bibinfo  {journal} {Phys. Rev. Lett.}\ }\textbf
  {\bibinfo {volume} {100}},\ \bibinfo {pages} {224101} (\bibinfo {year}
  {2008})}\BibitemShut {NoStop}%
\bibitem [{\citenamefont {Berman}\ and\ \citenamefont
  {Kolovski{\u{\i}}}(1992)}]{BerKol}%
  \BibitemOpen
  \bibfield  {author} {\bibinfo {author} {\bibfnamefont {G.~P.}\ \bibnamefont
  {Berman}}\ and\ \bibinfo {author} {\bibfnamefont {A.~R.}\ \bibnamefont
  {Kolovski{\u{\i}}}},\ }\href {\doibase 10.1070/PU1992v035n04ABEH002228}
  {\bibfield  {journal} {\bibinfo  {journal} {Sov. Phys. Usp.}\ }\textbf
  {\bibinfo {volume} {35}},\ \bibinfo {pages} {303} (\bibinfo {year}
  {1992})}\BibitemShut {NoStop}%
\bibitem [{\citenamefont {Kudo}\ and\ \citenamefont {Monteiro}(2008)}]{Kudo}%
  \BibitemOpen
  \bibfield  {author} {\bibinfo {author} {\bibfnamefont {K.}~\bibnamefont
  {Kudo}}\ and\ \bibinfo {author} {\bibfnamefont {T.~S.}\ \bibnamefont
  {Monteiro}},\ }\href {\doibase 10.1103/PhysRevE.77.055203} {\bibfield
  {journal} {\bibinfo  {journal} {Phys. Rev. E}\ }\textbf {\bibinfo {volume}
  {77}},\ \bibinfo {pages} {055203} (\bibinfo {year} {2008})}\BibitemShut
  {NoStop}%
\end{thebibliography}%

\end{document}